\documentclass[sn-mathphys,Numbered]{sn-jnl}

\usepackage{siunitx}
\usepackage{xcolor}
\usepackage{algorithm}%
\usepackage{algorithmicx}%
\usepackage{algpseudocode}%
\usepackage{amsmath,amssymb,amsfonts}%
\usepackage{amsthm}%
\usepackage[title]{appendix}%
\usepackage{booktabs}%
\usepackage{glossaries}
\usepackage{graphicx}%
\usepackage{listings}%
\usepackage{manyfoot}%
\usepackage{mathrsfs}%
\usepackage[version=3]{mhchem}
\usepackage{multirow}%
\usepackage{textcomp}%

\usepackage{gensymb}
\usepackage{comment}

\begin{document}

\title[Reversible phase transition in ReS$_2$]{Defect-assisted reversible phase transition in mono- and few-layer ReS$_2$}

\author[1]{\fnm{George} \sur{Zograf}}\equalcont{These authors contributed equally to this work.}

\author[1]{\fnm{Andrew B.} \sur{Yankovich}}\equalcont{These authors contributed equally to this work.}

\author[1]{\fnm{Bet$\textrm{{\"u}}$l} \sur{K$\textrm{{\"u}}$$\textrm{{\c{c}}}$$\textrm{{\"u}}$k$\textrm{{\"o}}$z}}

\author[1]{\fnm{Abhay V.} \sur{Agrawal}}

\author[1]{\fnm{Alexander Yu.} \sur{Polyakov}}

\author[2]{\fnm{Joachim} \sur{Ciers}}

\author[1]{\fnm{Fredrik} \sur{Eriksson}}

\author[2]{\fnm{$\textrm{\AA}$sa} \sur{Haglund}}

\author[1]{\fnm{Paul} \sur{Erhart}}

\author[1,3]{\fnm{Tomasz J.} \sur{Antosiewicz}}

\author[1]{\fnm{Eva} \sur{Olsson}}

\author*[1]{\fnm{Timur O.} \sur{Shegai}}\email{timurs@chalmers.se}


\affil[1]{\orgdiv{Department of Physics}, \orgname{Chalmers University of Technology},
\orgaddress{
\city{G$\textrm{{\"o}}$teborg}, \postcode{412 96}, 
\country{Sweden}}}

\affil[2]{\orgdiv{Department of Microtechnology and Nanoscience}, \orgname{Chalmers University of Technology}, 

\orgaddress{
\city{G$\textrm{{\"o}}$teborg}, \postcode{412 96},
\country{Sweden}}}

\affil[3]{\orgdiv{Faculty of Physics}, \orgname{University of Warsaw}, \orgaddress{\street{Pasteura 5}, \city{Warsaw}, \postcode{02-093}, \country{Poland}}}

\abstract{Transition metal dichalcogenide (TMD) materials have attracted substantial interest due to their remarkable excitonic, optical, electrical, and mechanical properties, which are highly dependent on their crystal structure. Controlling the crystal structure of these materials is essential for fine-tuning their performance, \textit{e.g.}, linear and nonlinear optical, as well as charge transport properties. While various phase-switching TMD materials, like molybdenum telluride (MoTe$_2$), are available, their transitions are often irreversible. Here, we investigate the mechanism of a light-induced reversible phase transition in mono- and bilayer flakes of rhenium disulfide (ReS$_2$). Our observations, based on scanning transmission electron microscopy, nonlinear spectroscopy, and density functional theory calculations, reveal a transition from the ground T$''$ (double distorted T) to the metastable H$'$ (distorted H) phase under femtosecond laser irradiation or influence of highly-energetic electrons. We show that the formation of sulfur vacancies facilitates this phenomenon. Our findings pave the way towards actively manipulating the crystal structure of ReS$_2$ and possibly its heterostructures.}

\keywords{reversible phase transition, transition metal dichalcogenide, ReS$_2$, second-harmonic generation}

\maketitle
\section*{Introduction}\label{sec1}

Over the past two decades, following the seminal work on graphene~\cite{novoselov2004electric}, there has been a substantial growth of interest in atomically thin two-dimensional (2D) materials. Among these, transition metal dichalcogenides (TMDs) have emerged as a particularly promising family due to their remarkable performance in various nanophotonic~\cite{mak2016photonics,trovatello2021optical}, nanoelectronic~\cite{radisavljevic2011single}, and other nanotechnology applications~\cite{manzeli20172d}. TMDs form a diverse material platform comprising transition metal (\textit{e.g.}, Mo, W, Re, Nb, Ta, \textit{etc}.) and chalcogenide elements (\textit{e.g.}, S, Se, Te). The flexibility to choose these constituent elements during synthesis enables the tunability of their optical~\cite{munkhbat2022optical} and electronic properties~\cite{wang2012electronics}. Furthermore, stacking atomically thin 2D materials in van der Waals heterostructures~\cite{geim2013van} offers control over their properties enabling various quantum phenomena, such as superconductivity in graphene-based structures~\cite{balents2020superconductivity,cao2018unconventional}, Moir{\'e} excitons and correlated states in TMDs~\cite{fang2019discovery,jin2019observation,alexeev2019resonantly,BreLinErh20,tang2020simulation,huang2021correlated}, topological excitons~\cite{wu2017topological}, advanced excitonic devices~\cite{ciarrocchi2022excitonic}, and exciton polaritons~\cite{zhang2021van,munkhbat2021tunable,dirnberger2023magneto}. Following the fabrication of large TMD flakes, achieved through methods like exfoliation~\cite{huang2020universal} or epitaxial single crystal growth~\cite{dumcenco2015large,li2021epitaxial}, further modification of material properties can be realized through nanopatterning, \textit{e.g.}, using electron beam lithography (EBL) and/or chemical etching. Nanopatterning of TMDs has enabled the production of high-quality factor (high-Q) nanophotonic structures~\cite{munkhbat2020transition,munkhbat2023nanostructured,weber2023intrinsic,zotev2023van}, exploration of higher order modes and anapole states~\cite{verre2019transition,tselikov2022transition,maciel2023probing,zograf2023combining}, and lasing from indirect bandgap semiconductor microdisks~\cite{sung2022room}. Expanding control over material properties beyond conventional fabrication methods will enable more precise engineering of future devices and deeper insights into the fundamental physics of TMD-related phenomena~\cite{ling2021all,ling2023deeply}.

The ability to actively and reversibly tune TMD properties post-fabrication is of high interest within the 2D material community~\cite{li2021phase}. Realizing this control to its full potential could expand the range of applications, including neuromorphic computing~\cite{chialvo2010emergent,markovic2020physics} as well as optically rewritable electronic circuits and memories~\cite{wuttig2007phase}. One approach for achieving active control over TMD electronic properties involves photo-induced modifications of the concentration of substitutional and interstitial point defects~\cite{seo2021reconfigurable}. Another approach leverages the substantial changes in properties associated with material polymorphism or phase transitions. For instance, entire TMD flakes can undergo phase transitions driven by electrostatic-doping~\cite{wang2017structural} or mechanical strain~\cite{voiry2013enhanced}. While these approaches show promise, they typically lack temporal stability and/or precise spatial control at the nanometer scale. Laser-induced phase patterning offers more precise spatial control but is often irreversible due to ablation~\cite{cho2015phase,guan2023femtosecond}, similar to thermal heating during the annealing process in 2D materials~\cite{yu2018high}. Nevertheless, promising routes for photo-induced phase switching exist~\cite{nasu2004photoinduced,yang2017structural}, such as the reversible transition of TaS$_2$ into a hidden stable quantum state, enabling drastic changes in conductivity~\cite{stojchevska2014ultrafast,vaskivskyi2024high}. Therefore, exploring new mechanisms for phase transitions in TMDs and understanding their associated property changes are crucial for achieving improved active control.

Rhenium disulfide (ReS$_2$) is a promising TMD material for optical, electronic, and gas-sensing applications~\cite{zhang2015res2,pradhan2015metal,rahman2017advent,xiang2019anomalous,gogna2020self,xiong2022nonvolatile}. Its features include being a direct bandgap semiconductor in a bulk van der Waals crystal~\cite{tongay2014monolayer} and its high optical anisotropy, making it ideal for directionally selective photonic applications~\cite{lin2015single,chenet2015plane,liu2016highly,munkhbat2022optical,ermolaev2024wandering}. Previous investigations have focused on ReS$_2$'s atomic structure and phases using density functional theory (DFT)~\cite{tongay2014monolayer} and scanning transmission electron microscopy (STEM)~\cite{zhou2020stacking}. Additionally, second-harmonic generation (SHG) studies have linked its activity to stacking and thickness variations~\cite{song2018extraordinary}. Notably, ReS$_2$ monolayers (ML) and bilayers (BL) have been demonstrated to exhibit a reversible phase transition under near-infrared femtosecond (fs) laser irradiation, characterized by a significant change in SHG activity and polarization-resolved SHG~\cite{kucukoz2022boosting}. It was hypothesized from the available data that ReS$_2$ transitions from the 1T$'$ phase into the 1H phase~\cite{kucukoz2022boosting}. Despite this promising initial work, questions remain regarding the involved transition phases and the mechanism enabling the transition.

Here, we investigate the phase transition in ReS$_2$ using a combination of electronic structure calculations, stability analysis, optical SHG microscopy, STEM imaging and diffraction, and electronic transport measurements. Our study results in several key observations: (i) femtosecond laser irradiation induces a phase transition in ReS$_2$ which converts back to the initial phase with time, (ii) the initial and the ``switched'' phase are identified as T$''$ and H$'$, respectively, with the latter being metastable (as opposed to its unstable H counterpart), and (iii) defects, particularly sulfur vacancies, play a crucial role in facilitating the 1T$''$-1H$'$ transition. Our investigation provides a deeper understanding of the possible phases of ML and BL ReS$_2$ and reveals the importance of vacancy defect formation on the mechanisms of phase switching.

\section*{Results \& Discussion}\label{sec2}

\begin{figure}
\centering
  \includegraphics[width=0.75\linewidth]{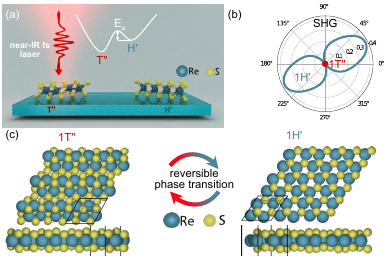}
  \caption{\textbf{Overview}. (a) Schematic of the physical process underpinning the laser-induced phase-transition. Femtosecond laser pulses in the near-IR range induce a phase transition from the T$''$ ground state of ReS$_2$ flakes into the H$'$ transient state manifested by a structural change. (b) SHG polarization-resolved intensity of 1T$''$ and 1H$'$ monolayer flakes calculated using DFT. (c) Structural changes in the lattice symmetry upon the phase transition between 1T$''$ and 1H$'$ monolayer flakes of ReS$_2$.}
  \label{fig:TOC}
\end{figure}

The main premise of our work is depicted in Figure~\ref{fig:TOC}. Under intense fs-laser illumination or influence of high-energy electrons, 1T$''$ ML and 2T$''$ BL flakes undergo a phase transition into what we hypothesize to be the 1H$'$ phase, a detailed description of which is provided in the Electronic band structure calculation section below. This phase transition is facilitated by generation of a sufficient amount of sulfur vacancies that allow the transition to occur by providing the necessary geometrical freedom. The H$'$ phase is metastable and is protected by an activation energy barrier ($E_a$), illustrated schematically in Figure~\ref{fig:TOC}a. The successful realization of this phase transition is visually evident in a significantly amplified SHG signal that strongly depends on the crystalline symmetry of the atomic structure, as shown in Figure~\ref{fig:TOC}b for ML ReS$_2$ using DFT calculations. The atomic rearrangement during the phase transition from 1T$''$ to 1H$'$ is illustrated in Figure~\ref{fig:TOC}c.

Our experimental investigation starts with the fabrication of ML and BL ReS$_2$ flakes on two different substrates \textit{via} mechanical exfoliation and dry transfer from a host crystal. Using this approach, we deposited flakes onto: (i) a silicon/silicon oxide (Si/SiO$_2$) substrate for optical and field-effect transistor (FET) transport studies, and (ii) a $\sim$ 20 nm thick silicon nitride (SiN) membrane TEM grid for STEM and electron diffraction studies. More details about the fabrication process are provided in the Methods section.

The thickness of the fabricated flakes was identified in three steps: optical imaging, SHG intensity, and atomic-force microscopy (AFM). Optical contrast in reflection microscopy provides an accurate estimation of the thickness of the flake~\cite{castellanos2010optical,anzai2019broad}. Owing to different symmetries upon stacking, the ReS$_2$ flakes demonstrate SHG activity depending on the number of layers~\cite{song2018extraordinary}.

\subsection*{Evidencing the phase transition in mono- and bilayer ReS$_2$ by electron microscopy}

\begin{figure}
\centering
  \includegraphics[width=0.9\linewidth]{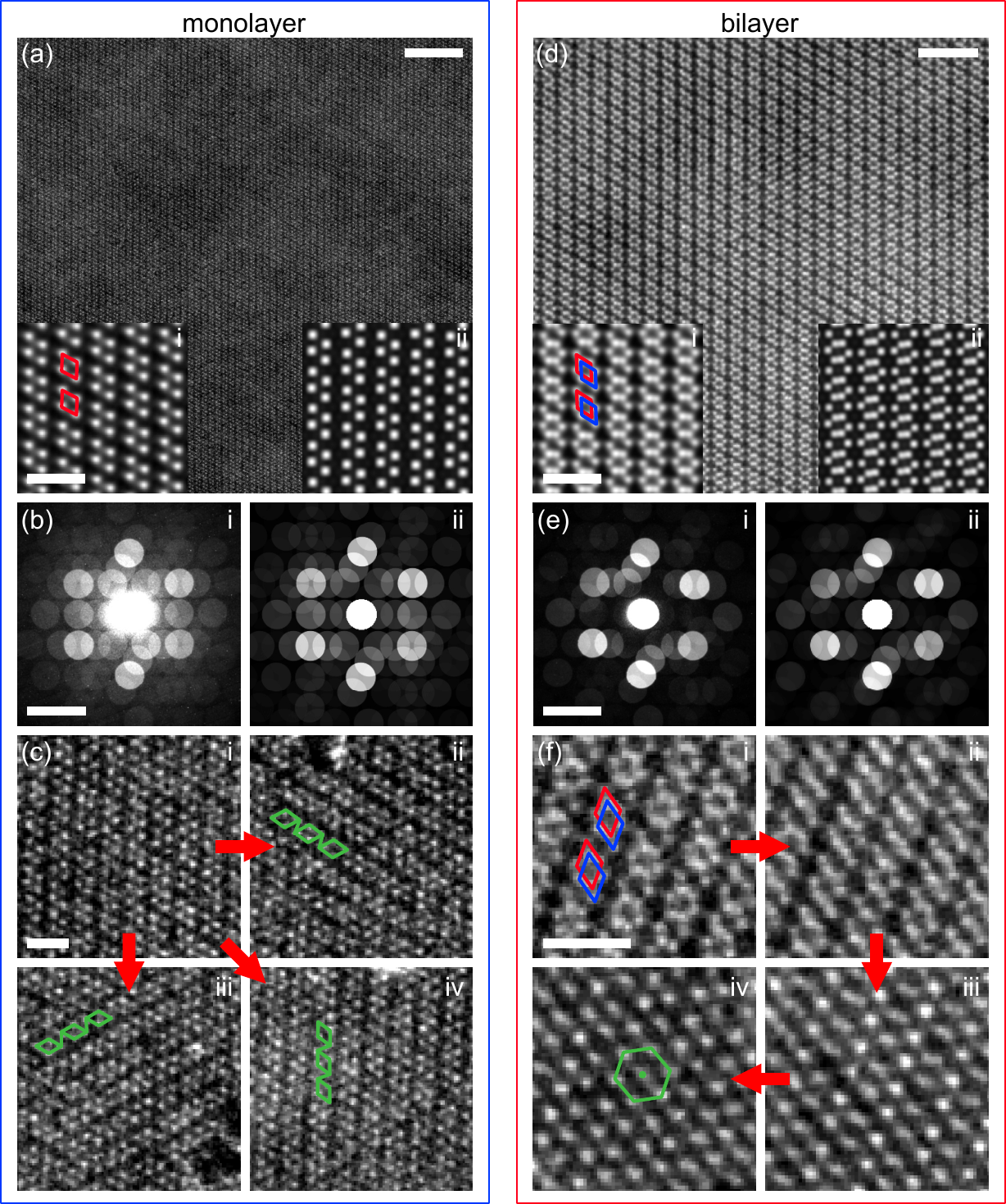}
  \caption{\textbf{ML and BL ReS$_2$ STEM observations}. (a) Experimental ADF STEM image of undamaged ML ReS$_2$. 
  The lower left inset image is a template-matched average image produced from the entire image in (a). The lower right inset image is a simulated ADF STEM image of the 1T$''$ phase.
  (b)(i) Experimental PACBED pattern of undamaged ML ReS$_2$. (b)(ii) Simulated PACBED pattern of the 1T$''$ phase.
  (c) Experimental ADF STEM images from different regions of electron-irradiated ML ReS$_2$. (c)(i) is from an undamaged 1T$''$ region. (c)(ii-iv) are from different damaged regions that exhibit a loss of the 1T$''$ phase but maintain a dominant distorted axis that rotates by 60 degrees around the out-of-plane c-axis to produce three possible in-plane orientations. The green markers represent the rearrangement of Re atoms upon phase transition.
  (d) Experimental ADF STEM image of undamaged BL ReS$_2$.
  The lower left inset image is a template-matched average image produced from the entire image in (d). The lower right inset image is a simulated ADF STEM image of the 2T$''$-e phase.
  (e)(i) Experimental PACBED pattern of undamaged BL ReS$_2$. (e)(ii) Simulated PACBED pattern of the 2T$''$e phase.
  (f) Experimental ADF STEM images extracted from an image series showing the atomic structure evolution of BL ReS$_2$ due to electron beam irradiation. (f)(i) shows the initial undamaged structure that is consistent with the 2T$''$ phase. (f)(ii-iv) shows three distinct structures that are observed in sequence during electron beam irradiation, which are followed by complete destruction of the 2D crystalline lattice. The red and blue markers in (a), (d), and (f) indicate the double distorted diamond structures of the T'' initial phase. The green markers in (c) and (f) indicate the single distorted Re chains in MLs and the hexagonal symmetry in BLs, respectively, upon phase transition.}
  \label{fig:TEM_Andy}
\end{figure}

To directly characterize the unperturbed starting phase of ML and BL ReS$_2$, as well as the evolution of their atomic structure during defect generation, we conducted experimental and simulated atomic resolution annular dark-field scanning transmission electron microscopy (ADF STEM) and position averaged convergent beam electron diffraction (PACBED) investigations. ADF STEM imaging enables one to directly visualize the positions of the Re atoms (see Methods for more details about the selection of STEM imaging modes and resulting image contrast) and, therefore, the atomic structure evolution. PACBED patterns provide a wealth of structural information about crystalline samples~\cite{lebeau2010position,ophus2019four}, such as the phase, by recording momentum-resolved measurements of the forward scattered electrons. PACBED enables the collection of this information from extremely small areas of the sample (down to a single unit cell or even atom~\cite{zhang2020atomic}), which can be controlled by selecting the sample area where the STEM probe is scanned. PACBED is particularly important to this work because it is sensitive to the electron scattering from both the Re and S atoms, and thus provides a S-sensitive phase characterization tool, in contrast to ADF STEM imaging.

We identify the structure of unperturbed ML ReS$_2$ using ADF STEM imaging and PACBED while considering 5 possible ML ReS$_2$ phases: 1H, 1T, 1T$'$, 1T$''$, and 1H$'$. The experimental atomic resolution ADF STEM image of unperturbed ML ReS$_2$ in Figure~\ref{fig:TEM_Andy}a and the associated template-averaged image in Figure~\ref{fig:TEM_Andy}(a,i) reveals the positions of the Re atoms and the presence of a double distorted structure that is indicated by the red diamonds. A simulated ADF STEM image of the 1T$''$ phase is shown in the lower right inset image. The similarity between the experimental (Figure~\ref{fig:TEM_Andy}(a,i)) and simulated (Figure~\ref{fig:TEM_Andy}(a,ii)) ADF STEM images, as well as the dissimilarity between the experimental ADF STEM image and simulated images of the other possible ML ReS$_2$ phases (see Figure S3), provide evidence that unperturbed ML ReS$_2$ has the 1T$''$ phase. In addition, the experimental PACBED pattern from ML ReS$_2$ (Figure~\ref{fig:TEM_Andy}(b,i)) is in best agreement with the simulated PACBED pattern from the 1T$''$ phase (Figure~\ref{fig:TEM_Andy}(b,ii) and Figure S3) providing further evidence that unperturbed ML ReS$_2$ has the 1T$''$ phase. 

Electron beam irradiation is known to introduce damage to TMD materials, primarily through a knock-on sputtering process that introduces chalcogenide vacancies~\cite{komsa2012two}. For the unperturbed phase measurements, special experimental design and analysis were conducted to ensure there were negligible effects from the electron beam irradiation on the phase identification (see Methods and Supplementary Information (SI) for more details). However, we also leveraged the chalcogenide vacancies production mechanism to design ADF STEM imaging experiments that resolve the atomic structure evolution of ML and BL ReS$_2$ during increasing S-vacancy concentrations. This is achieved by acquiring time-resolved ADF STEM image series with carefully selected electron beam energy, electron dose, and STEM image scan parameters (see Methods and SI).

During time-resolved ADF STEM imaging experiments on ML ReS$_2$ (see SI Movies 1 and 2 for example image series), electron irradiation damage occurs very rapidly, partially because experiments are limited to using a 200 keV electron beam due to sample constraints (see Methods and SI for more details). Similar behavior was observed in ML MoTe$_2$ during lower voltage TEM experiments~\cite{koster2023phase}. The image series shown in SI Movies 1 and 2 reveal that ML ReS$_2$ undergoes a complete destruction of its 2D crystallinity after a few tens of image frames (see SI for dose values). After a complete loss of the 2D crystalline structure, it appears that Re nanoclusters remain, providing evidence that the damage occurs through a sulfur sputtering and vacancy generation mechanism. Interestingly, just prior to the loss of the 2D crystallinity, ML ReS$_2$ undergoes a rearrangement of its atomic structure from a double distorted 1T$''$ phase to what appears like nano-sized regions of a \textit{single distorted} nonstoichiometric phase that can have three possible in-plane orientations. Figure~\ref{fig:TEM_Andy}(c,i-iv) shows extracted images from areas adjacent to and after SI Movie 2. Figure~\ref{fig:TEM_Andy}(c,i) shows the starting 1T$''$ phase, while Figure~\ref{fig:TEM_Andy}(c,ii-iv) show the three possible orientations of the resulting single distorted structure that are rotated by 60$\degree$ from each other. Figure S2 provides details on the image locations of Figure~\ref{fig:TEM_Andy}(c,i-iv) and a Fast Fourier transform analysis of these nano-sized regions that confirms a clear change in structure and a 60$\degree$ rotation between possible single distorted phase regions. Because the ADF STEM images are only sensitive to the Re atoms and not the S atoms, we can not unambiguously identify the resulting single distorted phase using this measurement. However, this investigation reveals ML ReS$_2$ transitions from the 1T$''$ phase to a single distorted nonstoichiometric phase that is consistent with 1H$'$ (Figure S3). Furthermore, the 1H$'$ phase is consistent with our SHG data and DFT calculations, as we show below.

We now use ADF STEM imaging and PACBED to identify the structure of unperturbed BL ReS$_2$. For this analysis, we consider 15 possible BL ReS$_2$ phases, including 2H, 2T, 2T$'$, and twelve different 2T$''$ phases (named 2T$''$c-n) that have previously been explored as possible stable phases~\cite{qiao2016polytypism}. The experimental atomic resolution ADF STEM image of unperturbed BL ReS$_2$ in \autoref{fig:TEM_Andy}d and the associated template-averaged image in \autoref{fig:TEM_Andy}(d,i) reveal the presence of a stacked double distorted structure that is indicated by the red and blue diamonds. A simulated ADF STEM image of the 2T$''$e phase is shown in \autoref{fig:TEM_Andy}(d,ii). The similarity between the experimental, \autoref{fig:TEM_Andy}(d,i), and simulated, \autoref{fig:TEM_Andy}(d,ii) ADF STEM images, as well as the dissimilarity between the experimental ADF STEM image and simulated images of the other possible BL ReS$_2$ phases (see Figures S7-S8), provides evidence that unperturbed BL ReS$_2$ has the 2T$''$e phase. In addition, the experimental PACBED pattern from BL ReS$_2$, \autoref{fig:TEM_Andy}(e,i), is in best agreement with the simulated PACBED pattern from the 2T$''$e phase (\autoref{fig:TEM_Andy}(e,ii), Figures S6 and S9) providing further evidence that unperturbed BL ReS$_2$ has the 2T$''$e phase. After considering all 12 possible phases, we can conclude that unperturbed BL ReS$_2$ has the 2T$''$ phase and most likely the 2T$''$e variant.

Time-resolved ADF STEM imaging experiments also provide evidence for electron beam-induced phase transitions in BL ReS$_2$. SI Movies 7-8 (see Methods and SI for data processing details and SI Movies 5-6 for unprocessed versions) are two consecutive image series of the same area which show that BL ReS$_2$ exhibits obvious structural changes during electron beam irradiation before eventually losing its 2D crystallinity. This is similar to what was observed for ML ReS$_2$, except BL ReS$_2$ requires a substantially higher electron dose for a complete loss of 2D crystallinity (see SI for dose values). Additionally, it undergoes what appears like three distinct atomic structure transitions before it loses its 2D crystallinity. \autoref{fig:TEM_Andy}(f,i-iv) are extracted images from SI Movies 7-8 at four different times (see SI). \autoref{fig:TEM_Andy}(f,i) shows the initial 2T$''$ phase characterized by the two stacked red and blue diamonds. After electron irradiation and the generation of S-vacancies, the structure transitions through 3 distinct phases shown in \autoref{fig:TEM_Andy}(f,ii-iv). \autoref{fig:TEM_Andy}(f,ii and iii) show the atomic structure transitions through what appears like two different single distorted phases. Following this, the final phase is characterized by the loss of the double and single distortion axes of the Re atoms and the presence of a hexagonal structure (see green hexagon in \autoref{fig:TEM_Andy}(f,iv)). Based on the atomic column contrast in \autoref{fig:TEM_Andy}(f,iv), each column appears to contain two Re atoms stacked in an A-A stacking configuration. Because the S atoms are invisible in these ADF STEM images and they are critical to identifying an H or T phase, it is not possible to unambiguously identify the transitional phases from this data (the T phase is, however, inconsistent with our SHG data, discussed below). Despite this, we can conclude the electron irradiation and the subsequent generation of S-vacancies causes BL ReS$_2$ to evolve from a starting double distorted 2T$''$ phase into nano-sized domains of two different nonstoichiometric single distorted phases before losing its Re distortion axes completely and transitioning into an A-A stacked hexagonal nonstoichiometric structure with a subsequent loss of its 2D crystallinity, leaving behind Re nanoclusters. Although we can not extract the S-vacancy concentration throughout this evolution, we expect the S-vacancy concentration to continuously increase during this process with the total electron dose, highlighting the critical role of defects in enabling the T$''$--H$'$ transition.

\subsection*{Light-induced phase transition} 

\begin{figure}
    \centering
    \includegraphics[width=0.6\linewidth]{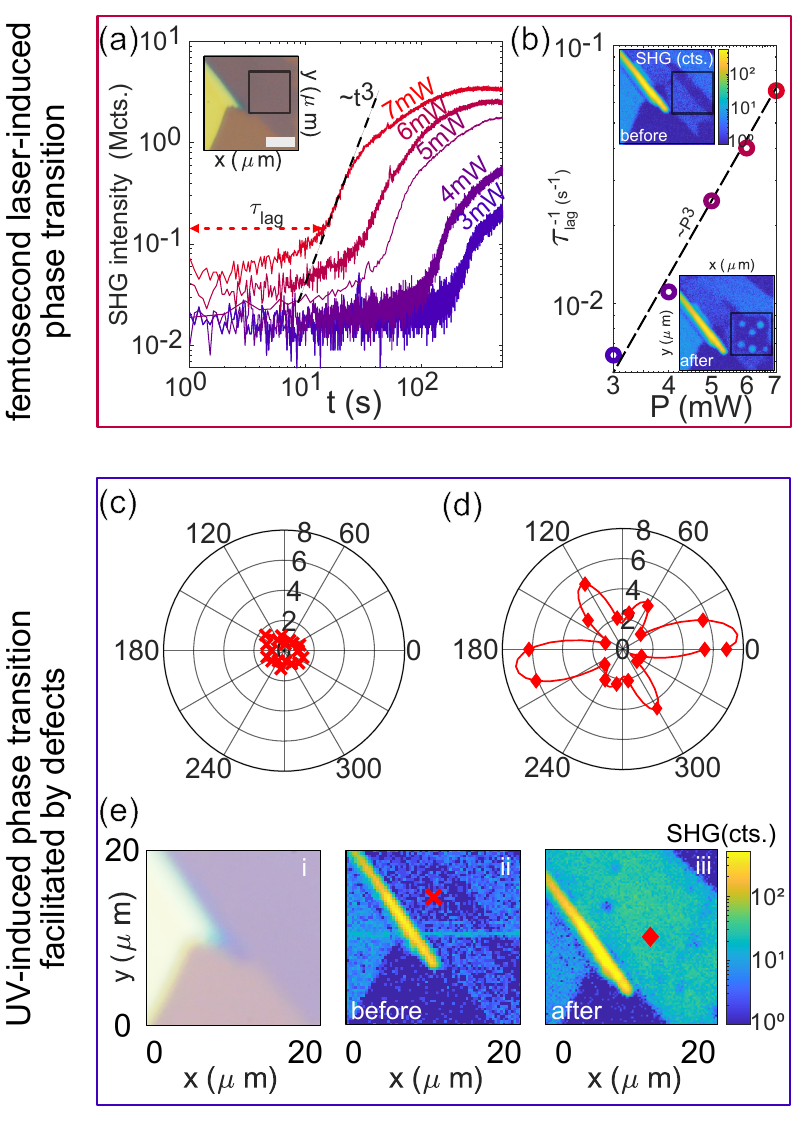}
    \caption{\textbf{Defect-induced phase transition in ReS$_2$ flakes.} (a) Time evolution of SHG counts for ReS$_2$ monolayer at 1040 nm fs-laser-induced point (static position of the laser beam) patterning process at different incident powers. $\tau_{\textrm{lag}}$ is the specific time of the SHG counts dramatic increase. The dashed line corresponds to the third power slope. The inset depicts the optical image of the flake, with the black square indicating the area of interest. The scale bar is 5 $\mu$m. (b) The rate of the initiation of the phase transition ($\tau^{-1}_{\textrm{lag}}$) as a function of incident power of the fs-laser at 1040 nm pump wavelength. The color of the experimental points corresponds to the same color as in the time evolution curve in (a). The dashed line corresponds to the third-power slope. Insets depict the SHG map before (upper) and after (lower) laser point patterning. (c) and (d) are polarization-resolved SHG polar plots obtained from respective areas of the panel (e) - red cross (e.ii - pristine) and red diamond (e.iii - exposed to moderate UV light). The flake after the formation of fs-laser-induced defects studied in (a,b) is subsequently exposed to UV light (e.iii). Panel (e) also shows an optical image of the flake (e.i). 
     }
    \label{fig:fs_laser}
\end{figure}

We now turn our attention to light-induced phase transition in ReS$_2$. In these experiments, we use a Si/SiO$_2$ substrate with a 285-nm-thick SiO$_2$ layer. We start our analysis by studying the time-dependent formation of the SHG-active 1H$'$ phase. \autoref{fig:fs_laser}a depicts the time evolution of the SHG intensity signal from different areas of the ReS$_2$ ML flake. The colors of the curves from blue to red demonstrate the increase of the incident 1040 nm wavelength fs-laser power. One can see similar features in every curve for different powers: (i) some time delay, or, as we introduce, the lag time $\tau_{\textrm{lag}}$ before the SHG intensity begins to substantially increase, (ii) saturation of the signal after reaching a certain level of counts. The analysis of the inverted lag time $\tau^{-1}_{\textrm{lag}}$ as a function of incident power, provides information about how soon the phase transition begins and how nonlinear this initiation process is. \autoref{fig:fs_laser}b depicts this information with a clear third-order dependence, indicating that the initiation of the phase transition of the ReS$_2$ ML on a given substrate is a third-order nonlinear optical process. This finding is of interest for patterning extended ReS$_2$ areas, as shown in comparison in the insets of Figure~\ref{fig:fs_laser}b before exposure (upper) and after exposure (lower). Such dependence of the lag time on the laser power may arise from two key processes -- multiple photon absorption of the flake required to induce the transition into the 1H$'$ phase and the formation of sulfur vacancies. The latter is in line with STEM and PACBED observations, discussed above. Additionally, it has been reported that ReS$_2$ MLs tend to host point defects more likely than, for example, MoS$_2$ as the covalent bond is rather soft in the 1T$''$ double-distorted phase~\cite{huang2018enhanced,horzum2014formation}.

Once the ML is heavily patterned by fs-laser (bottom inset \autoref{fig:fs_laser}b) and, therefore, hosts enough defects, one can induce a phase transition from 1T$''$ to 1H$'$ using an ultraviolet (UV) source at significantly smaller power compared to the fs-laser. Indeed, the polarization-resolved SHG plots in \autoref{fig:fs_laser}c,d before and after patterning manifest significant changes in the structure upon UV exposure. The pronounced 3 pairs of 2-fold symmetric SHG profiles with 60$\degree$ shift between each other in the patterned case indicate the formation of a different phase. The three 2-fold symmetries are in line with the STEM observations in~\autoref{fig:TEM_Andy}c and data provided in the SI, which indicate that upon phase transition ML ReS$_2$ forms distinct domains with the $b$-axes in each domain oriented 60$\degree$ with respect to each other. Individually, each domain yields an SHG spectrum with 2-fold symmetry, as calculated in~\autoref{fig:TOC}b, but there is no guarantee that in a given SHG measurement the surface area of the three mutually rotated 1H$'$ domains will be equal. Hence, in general, the SHG signal is expected to be characterized by three 2-fold symmetric profiles. Moreover, the intensity of the SHG from the ML in~\autoref{fig:fs_laser}e confirms the phase transition. The AFM data (see Figure S10) of the studied ReS$_2$ flake confirms its monolayer nature. 

We validate the defect-assisted nature of the light-induced phase transition in ReS$_2$ through various experimental techniques, including fs laser patterning, SHG polarization-resolved spectroscopy, UV-induced phase switching, and photoluminescence (PL) (the latter is discussed in SI, Figure S11). Furthermore, using the Si/SiO$_2$ substrate allows for conducting both optical and FET experiments and therefore gaining a deeper understanding of the electronic conductivity properties (see Figure S12). We note that the final phase of the ReS$_2$ can not be unambiguously identified solely from the SHG data, however, we conclude that ReS$_2$ switches into either H or H$'$ due to SHG intensity and polarization-resolved data pointing to the broken inversion symmetry of the lattice. Together with ADF STEM presented earlier and the subsequent discussion on DFT, we propose the hypothesis of T$''$--H$'$ phase transitions, as this scenario aligns consistently with all the experimental and theoretical methods used in our study as well as with observations from previously published work~\cite{kucukoz2022boosting}. Specifically, our observations indicate that a certain level of defects is necessary to facilitate the T$''$--H$'$ phase transition. These defects are induced either by fs laser irradiation or exposure to energetic electrons. The 1H$'$ phase is metastable due to the presence of an activation energy barrier, and once defects are introduced, the transition from the 1T$''$ phase to the 1H$'$ phase can be initiated by various external stimuli, such as UV light, fs laser pulses, or fast electrons. The formation of defects is irreversible, as evidenced by the irreversibility of the PL and FET measurements, which do not revert to their initial performance (see SI for details). However, the phase change in terms of SHG activity can be almost completely reversed at elevated temperatures, as discussed below.

\subsection*{Electronic structure calculations}

\begin{figure}[ht!]
    \centering
    \includegraphics[width=0.99\linewidth]{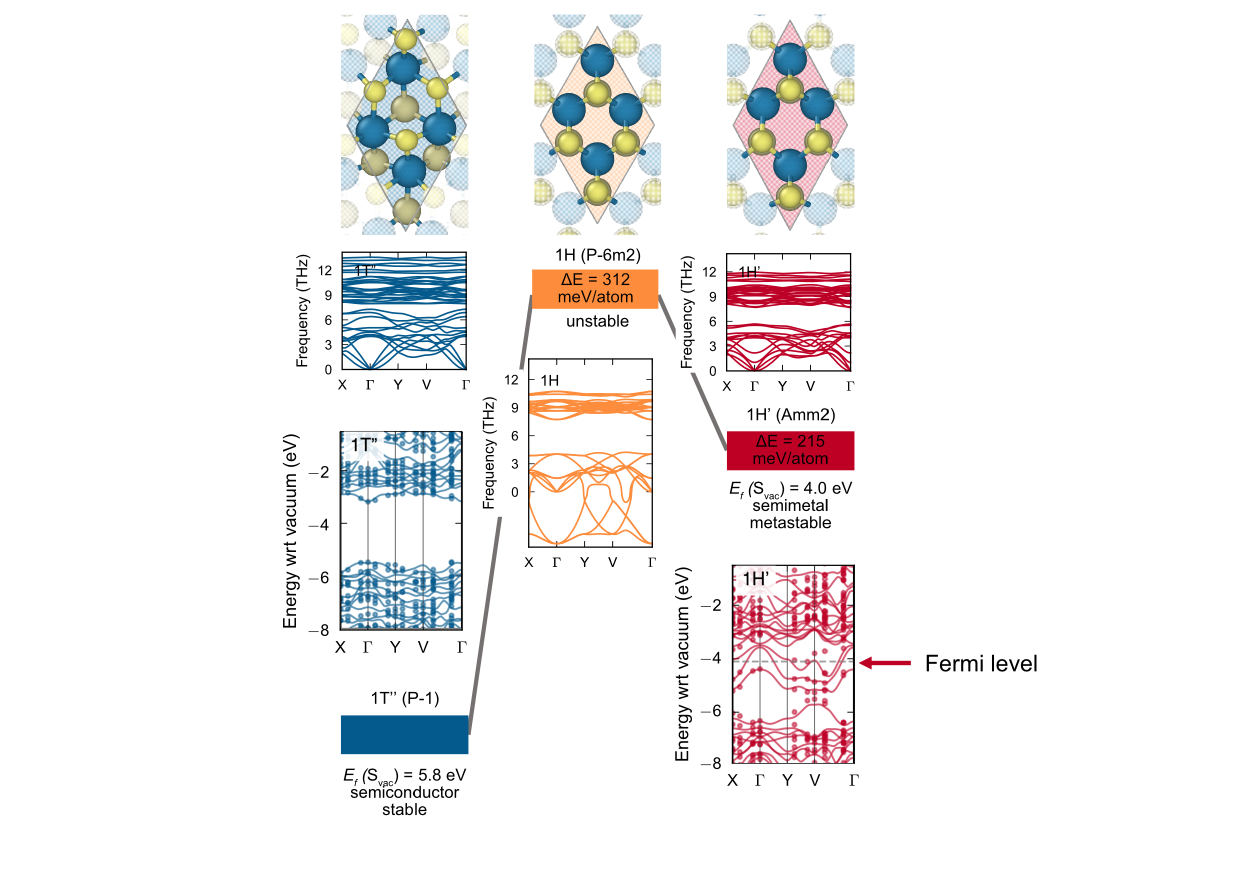}
    \caption{\textbf{DFT analysis of monolayer ReS$_2$ phases.} Left: stable 1T$''$ semiconducting phase. Middle: unstable 1H phase. Right: metastable 1H$'$ semimetal phase.}
    \label{fig:DFT}
\end{figure}

Our STEM and SHG data indicate a phase transition to a state with broken inversion symmetry. A reasonable first assumption is that the transition involves the standard undistorted H phase, similar to that of monolayers MoS$_2$, WS$_2$, and others~\cite{kucukoz2022boosting}. However, electronic structure calculations reveal that this phase is unstable in monolayer ReS$_2$~\cite{rahman2017advent}, see \autoref{fig:DFT}. This suggests that another phase, exhibiting broken inversion symmetry, likely underlies the observed SHG and STEM behaviors. STEM data further reveal that the ``switched'' phase is distorted (see~\autoref{fig:TEM_Andy}c), confirming its distinction from the 1H phase. Hence, we propose that this phase corresponds to 1H$'$ (the distorted 1H phase; space group $Amm2$, see \autoref{fig:TOC}c). Our numerical simulations, incorporating phononic and electronic band structure diagrams (see \autoref{fig:DFT}), indicate the stability of the 1H$'$ phase and its potential accessibility from 1T$''$ \textit{via} high-energy excitations utilizing photons (both multi- and single-photon processes are possible) or electrons (see Figure~S2). These calculations confirm that the 1H phase is unstable and therefore unlikely to be responsible for the observed behavior. Instead, ReS$_2$ tends to switch into a metastable 1H$'$ semimetal state, which aligns with our experimental observations in ADF STEM and SHG. To our knowledge, this is the first demonstration of the existence and stability of the 1H$'$ phase in ReS$_2$.

Our calculations suggest the involvement of sulfur vacancies in facilitating the T$''$--H$'$ phase transition. Furthermore, they also suggest the reduced energy required to form sulfur vacancies in the 1H$'$ phase compared to 1T$''$. The role of the sulfur defects consists in providing the necessary space for the transition to occur. Importantly, both the direct T$''$ to H$'$ and reverse H$'$ to T$''$ phase transitions require synchronized motion of large numbers of sulfur atoms, which can only occur when there is sufficient geometrical freedom to perform such motion. Additionally, molecular dynamics simulations demonstrate that the back-transition from 1H$'$ to 1T$''$ requires S-vacancies and represents a first-order phase transition involving nucleation and growth of one (switched) phase within the continuum of another (unswitched) phase (data not shown), confirming the role of sulfur vacancies. These predictions align with both light- and electron-induced 1T$''$--1H$'$ phase transitions observed in SHG and STEM measurements, as both require and indicate defect formation prior to the transition (for instance, see UV-induced phase change data,~\autoref{fig:fs_laser}c-e).

The SHG pattern of the proposed 1H$'$ phase can be simulated by using the $\chi^{(2)}$ polarizability tensor obtained from density functional theory (DFT) and yields a polarization-dependent profile with two distinct lobes (\autoref{fig:TOC}b). Our experimental SHG data suggests the existence of three possible orientation domains within the 1H$'$ phase (rotated with respect to each other by 60$\degree$). Their stochastic combination accounts for a 6-fold-like symmetric pattern (\autoref{fig:fs_laser}d), where SHG lobes vary in length, reflecting the prevalence of specific orientations in the final orientation domain composite (see STEM data in \autoref{fig:TEM_Andy}c).

Although numerical electronic structure calculations offer substantial support for the existence and relative stability of the 1H$'$ phase, alongside information about sulfur vacancies and second-order optical nonlinearities, they lack precision in estimating the activation energy barrier for the reverse H$'$ to T$''$ transition, which is crucial to characterize the stability of the H$'$ phase. Thus, in the subsequent section, we shift our focus to experimental observations of this back-switching process. The metastability of the H$'$ phase implies its eventual inevitable return to the stable T$''$ phase, with the switching rate following a temperature-dependent Arrhenius-like behavior, as we demonstrate below.

\begin{figure}[ht!]
    \centering
    \includegraphics[width=0.99\linewidth]{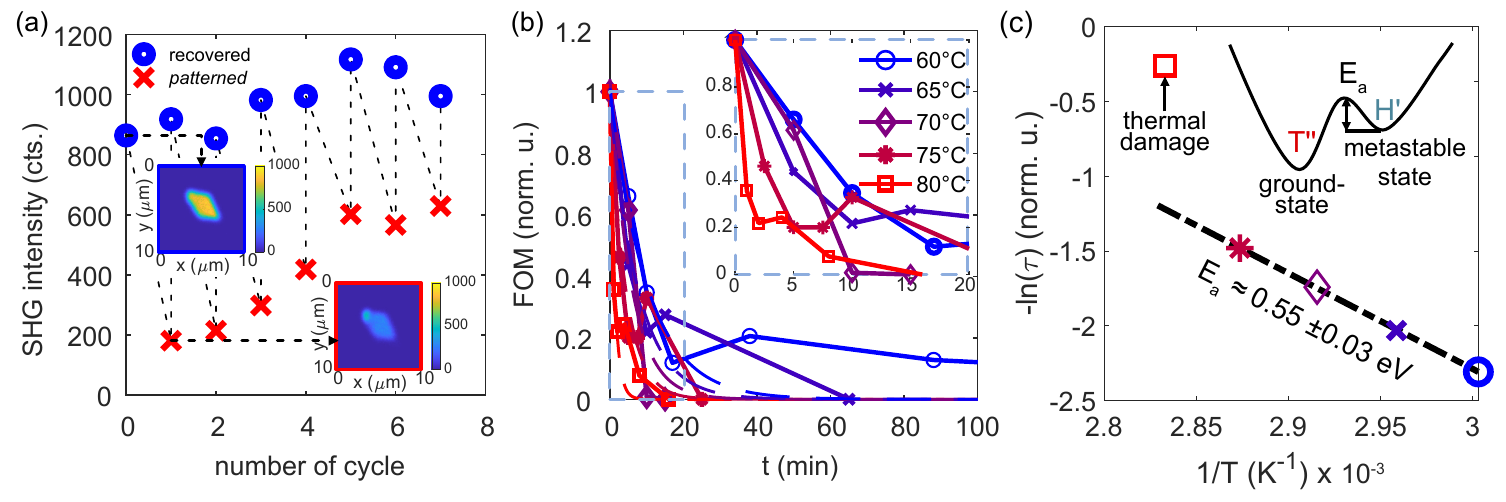}
    \caption{\textbf{Fs-laser-induced patterning mechanism.} (a) The average SHG intensity over the flake upon several patterning-recovery cycles of a bilayer ReS$_2$ flake. Blue circles correspond to the counts of SHG-active 2T$''$ of a bilayer, and the red crosses correspond to SHG counts of a 2H$'$ phase. The insets demonstrate the SHG maps of a bilayer. All the patterning cycles were carried out with the same conditions of the fs-laser. The dashed line links the measurements in chronological order. (b) Recovery FOM as a function of time for different temperature values that stimulated the recovery process. (c) FOM rate as an exponentially fitted recovery FOM function \textit{vs.} inverted temperature. The linear fitting reflects the energy activation barrier, $E_a$, between the phases. Colors and symbols correspond to the same colors and symbols as in (b). Inset is the sketch of the energy diagram corresponding to the T$''$--H$'$ phase transition.}
    \label{fig:E_a}
\end{figure}

\subsection*{Estimation of the activation energy barrier in bilayer ReS$_2$}

We use a pristine BL flake of ReS$_2$ to study the activation energy barrier height for the reverse transition from the H$'$ to the T$''$ phase. Unlike an ML (see~\autoref{fig:fs_laser}), the BL is highly SHG-active~\cite{song2018extraordinary,kucukoz2022boosting}, which is reflected in the upper inset with a blue frame of \autoref{fig:E_a}a. Upon fs-laser patterning of the area with specific step size (50 nm), exposure time (200 ms), and power (8 mW), the flake significantly reduces the counts of the SHG signal, indicating the phase transition from the T$''$ into H$'$ phase (see~\autoref{fig:E_a}a, lower inset with a red frame). The reduction of SHG signal in BL ReS$_2$ with laser patterning (as opposed to the enhancement of SHG in the ML case) is in line with previous observations~\cite{kucukoz2022boosting}. After some time, the sample recovers its initial SHG intensity counts, signaling the reverse transition to the pristine state. By applying the exposure conditions multiple times after each recovery, we repeated the patterning process 7 times for the bilayer to prove the reversibility of the phase transition, as shown in \autoref{fig:E_a}a. The blue circles depict the recovered (or pristine for the first blue circle) SHG-active T$''$ bilayer states, whereas the red crosses mark the patterned SHG-inactive H$'$ ones. Notably, we observe a clear trend that the difference in SHG signal between the blue and red points is reducing with the number of cycles. This could be due to incomplete recovery or substantial defect formation. Another important parameter is the time the BL ReS$_2$ flake resides in the metastable H$'$ phase. This time is temperature-dependent and hence allows to perform an Arrhenius-like study. The phase transition is reversible, therefore the activation energy barrier between ground T$''$ phase and H$'$ should not significantly exceed the thermal energy $k_BT$. With that in mind, we introduce the phase-transition recovery figure-of-merit (FOM), which relies on the SHG intensity as a measure of the phase transition. If the I$_{\textrm{rec}}$ and I$_{\textrm{patt}}$ are the SHG intensities of the recovered and patterned bilayer flake respectively, then FOM($t$,T) = $\frac{I_{\textrm{rec}}-I_{\textrm{T}}(t)}{I_{\textrm{rec}}-I_{\textrm{patt}}}$, where T is the temperature of the flake, $t$ is the time after the patterning took place and $I_{\textrm{T}}(t)$ -- the intensity of the SHG of the flake at a particular moment $t$ and at specific temperature T. For a range of temperatures from 60$\degree$C to 80$\degree$C with a 5$\degree$C step size, we performed a series of the measurements of the FOM($t$,T) as shown in \autoref{fig:E_a}b, for the same sample from \autoref{fig:E_a}a. FOM = 1 right at the moment of a freshly patterned flake (2H$'$), whereas FOM = 0 means that the sample recovered to its initial SHG intensity (2T$''$). The sample recovery curves at different temperatures can be fitted with a single exponential decay function, with a characteristic time scale $\tau$. The rate of the reverse phase transition is, therefore, $\tau^{-1}$. For every particular temperature, we plot the logarithm of $\tau^{-1}$. Furthermore, we fit the $\textrm{-ln}(\tau)$ versus 1/T using a linear function and obtain $E_a = 0.55 \pm 0.03$ eV with 95$\%$ confidence (see~\autoref{fig:E_a}c). The outlier experimental point corresponding to 80~$^{\circ}$C is not taken into account due to the observation of irreversible phase-switching behavior at higher temperatures, signaling the introduction of another process. Based on these experimental findings, we deduce that the stabilization energy barrier for the 2H$'$ phase is $\sim$ 0.55 eV. The phase transition diagram is schematically depicted in~\autoref{fig:E_a}c. Initially, the H$'$ phase can not form due to the insufficient number of defects. However, upon fs-laser-induced multiphoton absorption, combined with the simultaneous creation of S-vacancies, a new excited state relaxation pathway opens up, enabling the formation of the metastable H$'$ phase through the first-order phase transition, which involves nucleation and growth.

\section*{Conclusions} In this study, we present evidence of a reversible phase transition occurring in mono-, bi-, and trilayer ReS$_2$ (the latter is shown in Figure S11). The transition occurs from the T$''$ to the H$'$ phase at room temperature and ambient conditions. Through a combination of several experimental and theoretical methods, including electron microscopy, second-order nonlinear spectroscopy, photoluminescence, electronic transport, and DFT calculations, we identify that the transition is consistent with the metastable H$'$ (distorted H) phase. Notably, our data suggest a crucial role of sulfur vacancies in enabling this transition. Our analysis also confirms the ground state phase of the ReS$_2$ ML and BL flakes to be 1T$''$ and 2T$''$, respectively. Furthermore, we experimentally determine the activation energy barrier for the reverse H$'$ to T$''$ transition in BL ReS$_2$ to be approximately 0.55 eV. From a broader perspective, these results expand the knowledge of light- and electron-induced phase-switching materials and offer insights into the potential for optically programmable properties of TMD-based devices.

\section*{Methods}\label{secmet}

\textbf{Sample fabrication}. The ReS${_2}$ flakes were mechanically exfoliated from the high-quality bulk crystal (HQ Graphene) using scotch-tape with a subsequent transfer on polydimethylsiloxane stamps (PDMS) in a cleanroom (Nanofabrication laboratory at Chalmers within MyFab). For the STEM experiments, ReS$_2$ flakes were transferred to 20 nm thick SiN membrane TEM grids (simpore.com) using an all-dry transfer method~\cite{castellanos2014deterministic}. For the laser patterning and all the optical measurements, the flakes were transferred from the PDMS stamps to the glass slides.  For the electrical transport measurements, ReS$_2$ flakes were transferred to the 285-nm-thick SiO$_2$/Si highly $n$-doped substrates (Graphene Supermarket, USA) with pre-fabricated 20 nm Cr/200 nm Au back contact. To produce source and drain top contacts for ReS${_2}$ flakes, $\sim$ 300 nm layer of 950 PMMA A4 resist (MicroChem, USA) was applied by spin-coating. PMMA was patterned by EBL employing Raith EBPG 5200 (Germany) system operating at 100 kV accelerating voltage and 30 nA current. A dose of 1300 $\mu$C/cm$^{2}$ was applied, and 1:3 MBIK:IPA mixture was used as a developer. Then, 5~nm Cr/200 nm Au layer was evaporated by Lesker PVD 225 e-beam evaporation system (Kurt J. Lesker Company, Germany). Overnight lift-off in acetone was used to finalize the contact fabrication.

\textbf{Optical measurements and light-induced phase switching}. Second-harmonic generation (SHG) measurements were carried out using a tunable (690–1040 nm) Ti:sapphire femtosecond laser (MaiTai HP-Newport Spectra-Physics) with a $\sim$ 100 fs pulse duration and 80 MHz repetition rate. The light was polarized linearly with a broadband linear polarizer and focused through an objective (Nikon, 40$\times$, NA = 0.95). The positioning of the sample was controlled both in plane ($x - y$) and focus ($z$) by a piezoelectric stage. The signal was collected with an avalanche photo-diode (APD, IDQ, ID100 Visible Single-Photon Detector) for SHG mapping or a spectrometer (fiber-coupled Andor 500$i$, equipped with Newton 920 CCD camera) for SHG spectroscopy. The polarization-resolved SHG measurements were performed on the same setup using $\lambda/2$ plate to rotate the linear polarization of the excitation light.

PL measurements (Figure S11) were carried out using an inverted optical microscope (Nikon Eclipse 2000E) equipped with a high-NA oil immersion objective (Nikon, 60$\times$, NA = 1.49). The excitation 447 nm continuous-wave laser (Becker \& Hickl BDL-440-SMC Diode Laser) was coupled to an optical fiber and focused on the bottom side of the sample through the objective. A 488 nm long-pass filter is placed in the detection path to suppress the laser excitation. The setup allows for rotating the incident linear polarization and introducing an analyzer in the signal collection channel. The maximum input power is 10 mW.

The fs-laser induced phase switching experiments were performed using 1040 nm pump wavelength of the Ti:sapphire femtosecond laser (MaiTai HP-Newport Spectra-Physics) with a $\sim$ 100 fs pulse duration and 80 MHz repetition rate in a similar manner as described elsewhere~\cite{kucukoz2022boosting}. The laser beam was focused on a specific area of the flake with an approximate spot size of 1~$\mu$m using an objective (Nikon, 40$\times$, NA = 0.95). The piezoelectric stage allowed to position the sample in an $x-y$ plane with a 50~nm step every 200 ms to ensure a homogeneous phase switching over the flake's area. The pump powers used were between in the 3 -- 7 mW range, as shown in Figure~\ref{fig:fs_laser}.

UV light phase switching experiments were performed using the fourth harmonic of a Nd:YAG laser Teem Photonics SNU-20F-10C, emitting at 266 nm with 0.55 ns pulse length and 20.5 kHz repetition rate. The laser beam was defocused to keep the intensity low over the area of 3 $\times$ 5 mm$^2$. Averaged over time power was 2 mW, while the peak intensity reached $\sim$ 1.2 kW/cm$^2$.

\textbf{STEM measurements and calculations}. STEM imaging and PACBED experiments were carried out on a JEOL Mono NEO ARM 200F microscope. ADF STEM imaging was conducted using an accelerating voltage of 200 keV, a probe convergence half-angle of 26.7 mrad, an ADF detector collection angle range of 65-146 mrad, and a probe current of 44 picoampere (pA). In order to improve the quality of the ADF STEM images for the atomic structure analyses, a combination of non-rigid registration of image series~\cite{yankovich2014picometre} and template matching was used depending on the specific data set. Additional details about the acquisition parameters, electron dose, and processing details of each STEM image are provided in the SI. PACBED experiments~\cite{lebeau2010position} were conducted using an accelerating voltage of 80 keV, a probe convergence half-angle of 6.35 mrad, and a probe current of $<$3 pA. To resolve the ML and BL ReS${_2}$ PACBED patterns more clearly, careful corrections of the detector dark signal and SiN support signal were manually performed (more details provided in SI). ADF STEM and PACBED simulations were calculated using the frozen phonon multislice algorithm implemented in the GPU-accelerated software Prismatic 2.0~\cite{dacosta2021prismatic}. Additional details about the STEM experiments and simulations are provided in the SI.

\textbf{Electrical transport measurements}. 
FET measurements were conducted using two source channel -measurement units (B2901A, Keysight) connected through a General Purpose Interface Bus (GPIB) cable with a computer. The typical FET measurements were performed at room temperature. The back-gate voltage was systematically adjusted within the range of -10 V to 10 V, while simultaneously tuning the source-drain voltage between -1 V and 1 V.

\textbf{Electronic structure calculations}.
The energetics of different phases, their dynamic stability, defect formation energies, and electronic band structures were obtained \textit{via} DFT calculations that were carried out using the projector augmented-wave method \cite{Blo94} as implemented in the Vienna ab-initio simulation package \cite{KreHaf93, KreFur96}.
The exchange-correlation potential was represented using the van-der-Waals density functional with consistent exchange method \cite{DioRydSch04, BerHyl2014}.
All calculations were carried out using a plane-wave energy cutoff of 520~eV and for the calculation of the forces, a finer support grid was employed to improve their numerical accuracy.
The Brillouin zone was sampled using $\Gamma$-centered grid with a linear $\vec{k}$-point spacing of about 0.25~\AA$^{-1}$ and Gaussian smearing with a width of 0.1~eV.
This setup has been previously employed and tested for other layered materials \cite{LinErh16, EriFraLin23}.

For the ideal $1A$ and $Amm2$ structures the electronic band structures were furthermore computed using the $G_0W_0$ method \cite{HedLun70} on a $\Gamma$-centered \numproduct{8x8x1} grid.
These calculations employed 282 empty bands, thereby including states up to at least \qty{28}{\electronvolt} above the valence band edge ($1A$) or Fermi level ($Amm2$), and used 50 imaginary frequency and time grid points. 

Phonon dispersions were computed \textit{via} the \textsc{phonopy} package \cite{TogTan15} in combination with the \textsc{hiphive} package \cite{EriFraErh19} using force constants from DFT calculations.

The $\chi^{(2)}$ polarizability tensor was calculated using the method and tools outlined by Taghizadeh \textit{et al.} \cite{Taghizadeh2021}. Briefly, we calculated the nonlinear optical response using the DFT open-source GPAW package \cite{2005_PRB_71_035109_mortensen,2010_JPhysCondMat_22_253202_enkovaara}. We used the PBE exchange-correlation functional \cite{1996_PRL_77_3865_perdew},
the Kohn-Sham orbitals were expanded using a plane wave basis set with an energy cutoff of 500~eV, we used a $21\times21\times3$ Monkhorst-Pack grid for the $\mathbf{k}$-mesh and a grid spacing of less than 0.2~\AA.
The number of empty bands included in the sum over bands was set to three times the number of occupied bands. We used  Fermi-Dirac occupation number smearing with a factor of 20~meV and line-shape broadening of 50~meV. Time-reversal symmetry was imposed in order to reduce the $k$-integrals to half.

\section*{Data Availability} Data generated during the study is available upon reasonable request from the corresponding author. For Arxiv version the Supplementary file is not provided.

\section*{Conflicts of interest.} The authors declare no conflicts of interest.

\section*{Acknowledgements.} We thank B. Munkhbat (DTU) for the ReS$_2$ sample that has been studied in TEM. T.O.S. acknowledges funding from the Swedish Research Council (VR, research environment grant No. 2016-06059, VR project, grant No. 2017-04545 and grant No. 2022-03347), Chalmers Excellence Initiative Nano, 2D-TECH VINNOVA competence center (Ref. 2019-00068) and Olle Engkvist foundation (grant No. 211-0063). A.B.Y., E.O., P.E., and T.O.S. acknowledge funding from the Knut and Alice Wallenberg Foundation (KAW, under grant No. 2019.0140). A.B.Y. acknowledges funding from the Swedish Research Council (VR, under grant No. 2020-04986). T.J.A. acknowledges funding from the Polish National Science Center, project 2019/34/E/ST3/00359. This work was performed in part at Myfab Chalmers and the Chalmers Material Analysis Laboratory (CMAL). The DFT computations were enabled by resources provided by the Swedish National Infrastructure for Computing (SNIC) at NSC, C3SE, and PDC partially funded by the Swedish Research Council through grant agreement No. 2018-05973 and by the ICM-UW (Grant~\#G55-6).

\section*{Authors contribution.} G.Z., A.B.Y, B.K., E.O., and T.O.S. conceived the project idea. G.Z. and B.K. performed optical measurements. A.B.Y. performed and analyzed STEM and PACBED data. A.V.A. performed FET measurements. A.Y.P. fabricated the samples for FET experiments. J.C. and {\AA}.H. assisted with UV experiments. F.E., P.E., T.J.A. calculated dispersion with DFT. T.J.A. calculated the nonlinear response of the flakes. G.Z., A.B.Y, and T.O.S. wrote the manuscript with contributions from all co-authors. 




\begin{thebibliography}{90}
\ifx \bisbn   \undefined \def \bisbn  #1{ISBN #1}\fi
\ifx \binits  \undefined \def \binits#1{#1}\fi
\ifx \bauthor  \undefined \def \bauthor#1{#1}\fi
\ifx \batitle  \undefined \def \batitle#1{#1}\fi
\ifx \bjtitle  \undefined \def \bjtitle#1{#1}\fi
\ifx \bvolume  \undefined \def \bvolume#1{\textbf{#1}}\fi
\ifx \byear  \undefined \def \byear#1{#1}\fi
\ifx \bissue  \undefined \def \bissue#1{#1}\fi
\ifx \bfpage  \undefined \def \bfpage#1{#1}\fi
\ifx \blpage  \undefined \def \blpage #1{#1}\fi
\ifx \burl  \undefined \def \burl#1{\textsf{#1}}\fi
\ifx \doiurl  \undefined \def \doiurl#1{\url{https://doi.org/#1}}\fi
\ifx \betal  \undefined \def \betal{\textit{et al.}}\fi
\ifx \binstitute  \undefined \def \binstitute#1{#1}\fi
\ifx \binstitutionaled  \undefined \def \binstitutionaled#1{#1}\fi
\ifx \bctitle  \undefined \def \bctitle#1{#1}\fi
\ifx \beditor  \undefined \def \beditor#1{#1}\fi
\ifx \bpublisher  \undefined \def \bpublisher#1{#1}\fi
\ifx \bbtitle  \undefined \def \bbtitle#1{#1}\fi
\ifx \bedition  \undefined \def \bedition#1{#1}\fi
\ifx \bseriesno  \undefined \def \bseriesno#1{#1}\fi
\ifx \blocation  \undefined \def \blocation#1{#1}\fi
\ifx \bsertitle  \undefined \def \bsertitle#1{#1}\fi
\ifx \bsnm \undefined \def \bsnm#1{#1}\fi
\ifx \bsuffix \undefined \def \bsuffix#1{#1}\fi
\ifx \bparticle \undefined \def \bparticle#1{#1}\fi
\ifx \barticle \undefined \def \barticle#1{#1}\fi
\bibcommenthead
\ifx \bconfdate \undefined \def \bconfdate #1{#1}\fi
\ifx \botherref \undefined \def \botherref #1{#1}\fi
\ifx \url \undefined \def \url#1{\textsf{#1}}\fi
\ifx \bchapter \undefined \def \bchapter#1{#1}\fi
\ifx \bbook \undefined \def \bbook#1{#1}\fi
\ifx \bcomment \undefined \def \bcomment#1{#1}\fi
\ifx \oauthor \undefined \def \oauthor#1{#1}\fi
\ifx \citeauthoryear \undefined \def \citeauthoryear#1{#1}\fi
\ifx \endbibitem  \undefined \def \endbibitem {}\fi
\ifx \bconflocation  \undefined \def \bconflocation#1{#1}\fi
\ifx \arxivurl  \undefined \def \arxivurl#1{\textsf{#1}}\fi
\csname PreBibitemsHook\endcsname

\bibitem{novoselov2004electric}
\begin{barticle}
\bauthor{\bsnm{Novoselov}, \binits{K.S.}},
\bauthor{\bsnm{Geim}, \binits{A.K.}},
\bauthor{\bsnm{Morozov}, \binits{S.V.}},
\bauthor{\bsnm{Jiang}, \binits{D.-e.}},
\bauthor{\bsnm{Zhang}, \binits{Y.}},
\bauthor{\bsnm{Dubonos}, \binits{S.V.}},
\bauthor{\bsnm{Grigorieva}, \binits{I.V.}},
\bauthor{\bsnm{Firsov}, \binits{A.A.}}:
\batitle{Electric field effect in atomically thin carbon films}.
\bjtitle{Science}
\bvolume{306}(\bissue{5696}),
\bfpage{666}--\blpage{669}
(\byear{2004})
\end{barticle}
\endbibitem

\bibitem{mak2016photonics}
\begin{barticle}
\bauthor{\bsnm{Mak}, \binits{K.F.}},
\bauthor{\bsnm{Shan}, \binits{J.}}:
\batitle{Photonics and optoelectronics of 2{D} semiconductor transition metal
  dichalcogenides}.
\bjtitle{Nat. Photon.}
\bvolume{10}(\bissue{4}),
\bfpage{216}--\blpage{226}
(\byear{2016})
\end{barticle}
\endbibitem

\bibitem{trovatello2021optical}
\begin{barticle}
\bauthor{\bsnm{Trovatello}, \binits{C.}},
\bauthor{\bsnm{Marini}, \binits{A.}},
\bauthor{\bsnm{Xu}, \binits{X.}},
\bauthor{\bsnm{Lee}, \binits{C.}},
\bauthor{\bsnm{Liu}, \binits{F.}},
\bauthor{\bsnm{Curreli}, \binits{N.}},
\bauthor{\bsnm{Manzoni}, \binits{C.}},
\bauthor{\bsnm{Dal~Conte}, \binits{S.}},
\bauthor{\bsnm{Yao}, \binits{K.}},
\bauthor{\bsnm{Ciattoni}, \binits{A.}}, \betal:
\batitle{Optical parametric amplification by monolayer transition metal
  dichalcogenides}.
\bjtitle{Nat. Photon.}
\bvolume{15}(\bissue{1}),
\bfpage{6}--\blpage{10}
(\byear{2021})
\end{barticle}
\endbibitem

\bibitem{radisavljevic2011single}
\begin{barticle}
\bauthor{\bsnm{Radisavljevic}, \binits{B.}},
\bauthor{\bsnm{Radenovic}, \binits{A.}},
\bauthor{\bsnm{Brivio}, \binits{J.}},
\bauthor{\bsnm{Giacometti}, \binits{V.}},
\bauthor{\bsnm{Kis}, \binits{A.}}:
\batitle{Single-layer {MoS$_2$} transistors}.
\bjtitle{Nat. Nanotechnol.}
\bvolume{6}(\bissue{3}),
\bfpage{147}--\blpage{150}
(\byear{2011})
\end{barticle}
\endbibitem

\bibitem{manzeli20172d}
\begin{barticle}
\bauthor{\bsnm{Manzeli}, \binits{S.}},
\bauthor{\bsnm{Ovchinnikov}, \binits{D.}},
\bauthor{\bsnm{Pasquier}, \binits{D.}},
\bauthor{\bsnm{Yazyev}, \binits{O.V.}},
\bauthor{\bsnm{Kis}, \binits{A.}}:
\batitle{2{D} transition metal dichalcogenides}.
\bjtitle{Nat. Rev. Mater.}
\bvolume{2}(\bissue{8}),
\bfpage{1}--\blpage{15}
(\byear{2017})
\end{barticle}
\endbibitem

\bibitem{munkhbat2022optical}
\begin{barticle}
\bauthor{\bsnm{Munkhbat}, \binits{B.}},
\bauthor{\bsnm{Wr{\'o}bel}, \binits{P.}},
\bauthor{\bsnm{Antosiewicz}, \binits{T.J.}},
\bauthor{\bsnm{Shegai}, \binits{T.O.}}:
\batitle{Optical constants of several multilayer transition metal
  dichalcogenides measured by spectroscopic ellipsometry in the 300--1700 nm
  range: High index, anisotropy, and hyperbolicity}.
\bjtitle{ACS Photonics}
\bvolume{9}(\bissue{7}),
\bfpage{2398}--\blpage{2407}
(\byear{2022})
\end{barticle}
\endbibitem

\bibitem{wang2012electronics}
\begin{barticle}
\bauthor{\bsnm{Wang}, \binits{Q.H.}},
\bauthor{\bsnm{Kalantar-Zadeh}, \binits{K.}},
\bauthor{\bsnm{Kis}, \binits{A.}},
\bauthor{\bsnm{Coleman}, \binits{J.N.}},
\bauthor{\bsnm{Strano}, \binits{M.S.}}:
\batitle{Electronics and optoelectronics of two-dimensional transition metal
  dichalcogenides}.
\bjtitle{Nat. Nanotechnol.}
\bvolume{7}(\bissue{11}),
\bfpage{699}--\blpage{712}
(\byear{2012})
\end{barticle}
\endbibitem

\bibitem{geim2013van}
\begin{barticle}
\bauthor{\bsnm{Geim}, \binits{A.K.}},
\bauthor{\bsnm{Grigorieva}, \binits{I.V.}}:
\batitle{Van der {W}aals heterostructures}.
\bjtitle{Nature}
\bvolume{499}(\bissue{7459}),
\bfpage{419}--\blpage{425}
(\byear{2013})
\end{barticle}
\endbibitem

\bibitem{balents2020superconductivity}
\begin{barticle}
\bauthor{\bsnm{Balents}, \binits{L.}},
\bauthor{\bsnm{Dean}, \binits{C.R.}},
\bauthor{\bsnm{Efetov}, \binits{D.K.}},
\bauthor{\bsnm{Young}, \binits{A.F.}}:
\batitle{Superconductivity and strong correlations in moir{\'e} flat bands}.
\bjtitle{Nat. Phys.}
\bvolume{16}(\bissue{7}),
\bfpage{725}--\blpage{733}
(\byear{2020})
\end{barticle}
\endbibitem

\bibitem{cao2018unconventional}
\begin{barticle}
\bauthor{\bsnm{Cao}, \binits{Y.}},
\bauthor{\bsnm{Fatemi}, \binits{V.}},
\bauthor{\bsnm{Fang}, \binits{S.}},
\bauthor{\bsnm{Watanabe}, \binits{K.}},
\bauthor{\bsnm{Taniguchi}, \binits{T.}},
\bauthor{\bsnm{Kaxiras}, \binits{E.}},
\bauthor{\bsnm{Jarillo-Herrero}, \binits{P.}}:
\batitle{Unconventional superconductivity in magic-angle graphene
  superlattices}.
\bjtitle{Nature}
\bvolume{556}(\bissue{7699}),
\bfpage{43}--\blpage{50}
(\byear{2018})
\end{barticle}
\endbibitem

\bibitem{fang2019discovery}
\begin{barticle}
\bauthor{\bsnm{Fang}, \binits{Y.}},
\bauthor{\bsnm{Pan}, \binits{J.}},
\bauthor{\bsnm{Zhang}, \binits{D.}},
\bauthor{\bsnm{Wang}, \binits{D.}},
\bauthor{\bsnm{Hirose}, \binits{H.T.}},
\bauthor{\bsnm{Terashima}, \binits{T.}},
\bauthor{\bsnm{Uji}, \binits{S.}},
\bauthor{\bsnm{Yuan}, \binits{Y.}},
\bauthor{\bsnm{Li}, \binits{W.}},
\bauthor{\bsnm{Tian}, \binits{Z.}}, \betal:
\batitle{Discovery of superconductivity in {2M WS$_2$} with possible
  topological surface states}.
\bjtitle{Adv. Mater.}
\bvolume{31}(\bissue{30}),
\bfpage{1901942}
(\byear{2019})
\end{barticle}
\endbibitem

\bibitem{jin2019observation}
\begin{barticle}
\bauthor{\bsnm{Jin}, \binits{C.}},
\bauthor{\bsnm{Regan}, \binits{E.C.}},
\bauthor{\bsnm{Yan}, \binits{A.}},
\bauthor{\bsnm{Iqbal Bakti~Utama}, \binits{M.}},
\bauthor{\bsnm{Wang}, \binits{D.}},
\bauthor{\bsnm{Zhao}, \binits{S.}},
\bauthor{\bsnm{Qin}, \binits{Y.}},
\bauthor{\bsnm{Yang}, \binits{S.}},
\bauthor{\bsnm{Zheng}, \binits{Z.}},
\bauthor{\bsnm{Shi}, \binits{S.}}, \betal:
\batitle{Observation of moir{\'e} excitons in {WSe$_2$/WS$_2$} heterostructure
  superlattices}.
\bjtitle{Nature}
\bvolume{567}(\bissue{7746}),
\bfpage{76}--\blpage{80}
(\byear{2019})
\end{barticle}
\endbibitem

\bibitem{alexeev2019resonantly}
\begin{barticle}
\bauthor{\bsnm{Alexeev}, \binits{E.M.}},
\bauthor{\bsnm{Ruiz-Tijerina}, \binits{D.A.}},
\bauthor{\bsnm{Danovich}, \binits{M.}},
\bauthor{\bsnm{Hamer}, \binits{M.J.}},
\bauthor{\bsnm{Terry}, \binits{D.J.}},
\bauthor{\bsnm{Nayak}, \binits{P.K.}},
\bauthor{\bsnm{Ahn}, \binits{S.}},
\bauthor{\bsnm{Pak}, \binits{S.}},
\bauthor{\bsnm{Lee}, \binits{J.}},
\bauthor{\bsnm{Sohn}, \binits{J.I.}}, \betal:
\batitle{Resonantly hybridized excitons in moir{\'e} superlattices in van der
  {W}aals heterostructures}.
\bjtitle{Nature}
\bvolume{567}(\bissue{7746}),
\bfpage{81}--\blpage{86}
(\byear{2019})
\end{barticle}
\endbibitem

\bibitem{BreLinErh20}
\begin{barticle}
\bauthor{\bsnm{Brem}, \binits{S.}},
\bauthor{\bsnm{Linder{\"a}lv}, \binits{C.}},
\bauthor{\bsnm{Erhart}, \binits{P.}},
\bauthor{\bsnm{Malic}, \binits{E.}}:
\batitle{Tunable phases of {M}oiré excitons in van der {W}aals
  heterostructures}.
\bjtitle{Nano Lett.}
\bvolume{20},
\bfpage{8534}
(\byear{2020})
\end{barticle}
\endbibitem

\bibitem{tang2020simulation}
\begin{barticle}
\bauthor{\bsnm{Tang}, \binits{Y.}},
\bauthor{\bsnm{Li}, \binits{L.}},
\bauthor{\bsnm{Li}, \binits{T.}},
\bauthor{\bsnm{Xu}, \binits{Y.}},
\bauthor{\bsnm{Liu}, \binits{S.}},
\bauthor{\bsnm{Barmak}, \binits{K.}},
\bauthor{\bsnm{Watanabe}, \binits{K.}},
\bauthor{\bsnm{Taniguchi}, \binits{T.}},
\bauthor{\bsnm{MacDonald}, \binits{A.H.}},
\bauthor{\bsnm{Shan}, \binits{J.}}, \betal:
\batitle{Simulation of hubbard model physics in {WSe$_2$/WS$_2$} moir{\'e}
  superlattices}.
\bjtitle{Nature}
\bvolume{579}(\bissue{7799}),
\bfpage{353}--\blpage{358}
(\byear{2020})
\end{barticle}
\endbibitem

\bibitem{huang2021correlated}
\begin{barticle}
\bauthor{\bsnm{Huang}, \binits{X.}},
\bauthor{\bsnm{Wang}, \binits{T.}},
\bauthor{\bsnm{Miao}, \binits{S.}},
\bauthor{\bsnm{Wang}, \binits{C.}},
\bauthor{\bsnm{Li}, \binits{Z.}},
\bauthor{\bsnm{Lian}, \binits{Z.}},
\bauthor{\bsnm{Taniguchi}, \binits{T.}},
\bauthor{\bsnm{Watanabe}, \binits{K.}},
\bauthor{\bsnm{Okamoto}, \binits{S.}},
\bauthor{\bsnm{Xiao}, \binits{D.}}, \betal:
\batitle{Correlated insulating states at fractional fillings of the
  {WS$_2$/WS$_2$} moir{\'e} lattice}.
\bjtitle{Nat. Phys.}
\bvolume{17}(\bissue{6}),
\bfpage{715}--\blpage{719}
(\byear{2021})
\end{barticle}
\endbibitem

\bibitem{wu2017topological}
\begin{barticle}
\bauthor{\bsnm{Wu}, \binits{F.}},
\bauthor{\bsnm{Lovorn}, \binits{T.}},
\bauthor{\bsnm{MacDonald}, \binits{A.H.}}:
\batitle{Topological exciton bands in moir{\'e} heterojunctions}.
\bjtitle{Phys. Rev. Lett.}
\bvolume{118}(\bissue{14}),
\bfpage{147401}
(\byear{2017})
\end{barticle}
\endbibitem

\bibitem{ciarrocchi2022excitonic}
\begin{barticle}
\bauthor{\bsnm{Ciarrocchi}, \binits{A.}},
\bauthor{\bsnm{Tagarelli}, \binits{F.}},
\bauthor{\bsnm{Avsar}, \binits{A.}},
\bauthor{\bsnm{Kis}, \binits{A.}}:
\batitle{Excitonic devices with van der {W}aals heterostructures: valleytronics
  meets twistronics}.
\bjtitle{Nat. Rev. Mater.}
\bvolume{7}(\bissue{6}),
\bfpage{449}--\blpage{464}
(\byear{2022})
\end{barticle}
\endbibitem

\bibitem{zhang2021van}
\begin{barticle}
\bauthor{\bsnm{Zhang}, \binits{L.}},
\bauthor{\bsnm{Wu}, \binits{F.}},
\bauthor{\bsnm{Hou}, \binits{S.}},
\bauthor{\bsnm{Zhang}, \binits{Z.}},
\bauthor{\bsnm{Chou}, \binits{Y.-H.}},
\bauthor{\bsnm{Watanabe}, \binits{K.}},
\bauthor{\bsnm{Taniguchi}, \binits{T.}},
\bauthor{\bsnm{Forrest}, \binits{S.R.}},
\bauthor{\bsnm{Deng}, \binits{H.}}:
\batitle{Van der {Wa}als heterostructure polaritons with moir{\'e}-induced
  nonlinearity}.
\bjtitle{Nature}
\bvolume{591}(\bissue{7848}),
\bfpage{61}--\blpage{65}
(\byear{2021})
\end{barticle}
\endbibitem

\bibitem{munkhbat2021tunable}
\begin{barticle}
\bauthor{\bsnm{Munkhbat}, \binits{B.}},
\bauthor{\bsnm{Canales}, \binits{A.}},
\bauthor{\bsnm{K{\"u}{\c{c}}{\"u}k{\"o}z}, \binits{B.}},
\bauthor{\bsnm{Baranov}, \binits{D.G.}},
\bauthor{\bsnm{Shegai}, \binits{T.O.}}:
\batitle{Tunable self-assembled {C}asimir microcavities and polaritons}.
\bjtitle{Nature}
\bvolume{597}(\bissue{7875}),
\bfpage{214}--\blpage{219}
(\byear{2021})
\end{barticle}
\endbibitem

\bibitem{dirnberger2023magneto}
\begin{barticle}
\bauthor{\bsnm{Dirnberger}, \binits{F.}},
\bauthor{\bsnm{Quan}, \binits{J.}},
\bauthor{\bsnm{Bushati}, \binits{R.}},
\bauthor{\bsnm{Diederich}, \binits{G.M.}},
\bauthor{\bsnm{Florian}, \binits{M.}},
\bauthor{\bsnm{Klein}, \binits{J.}},
\bauthor{\bsnm{Mosina}, \binits{K.}},
\bauthor{\bsnm{Sofer}, \binits{Z.}},
\bauthor{\bsnm{Xu}, \binits{X.}},
\bauthor{\bsnm{Kamra}, \binits{A.}}, \betal:
\batitle{Magneto-optics in a van der {W}aals magnet tuned by self-hybridized
  polaritons}.
\bjtitle{Nature}
\bvolume{620}(\bissue{7974}),
\bfpage{533}--\blpage{537}
(\byear{2023})
\end{barticle}
\endbibitem

\bibitem{huang2020universal}
\begin{barticle}
\bauthor{\bsnm{Huang}, \binits{Y.}},
\bauthor{\bsnm{Pan}, \binits{Y.-H.}},
\bauthor{\bsnm{Yang}, \binits{R.}},
\bauthor{\bsnm{Bao}, \binits{L.-H.}},
\bauthor{\bsnm{Meng}, \binits{L.}},
\bauthor{\bsnm{Luo}, \binits{H.-L.}},
\bauthor{\bsnm{Cai}, \binits{Y.-Q.}},
\bauthor{\bsnm{Liu}, \binits{G.-D.}},
\bauthor{\bsnm{Zhao}, \binits{W.-J.}},
\bauthor{\bsnm{Zhou}, \binits{Z.}}, \betal:
\batitle{Universal mechanical exfoliation of large-area 2{D} crystals}.
\bjtitle{Nat. Commun.}
\bvolume{11}(\bissue{1}),
\bfpage{2453}
(\byear{2020})
\end{barticle}
\endbibitem

\bibitem{dumcenco2015large}
\begin{barticle}
\bauthor{\bsnm{Dumcenco}, \binits{D.}},
\bauthor{\bsnm{Ovchinnikov}, \binits{D.}},
\bauthor{\bsnm{Marinov}, \binits{K.}},
\bauthor{\bsnm{Lazic}, \binits{P.}},
\bauthor{\bsnm{Gibertini}, \binits{M.}},
\bauthor{\bsnm{Marzari}, \binits{N.}},
\bauthor{\bsnm{Sanchez}, \binits{O.L.}},
\bauthor{\bsnm{Kung}, \binits{Y.-C.}},
\bauthor{\bsnm{Krasnozhon}, \binits{D.}},
\bauthor{\bsnm{Chen}, \binits{M.-W.}}, \betal:
\batitle{Large-area epitaxial monolayer {MoS$_2$}}.
\bjtitle{ACS Nano}
\bvolume{9}(\bissue{4}),
\bfpage{4611}--\blpage{4620}
(\byear{2015})
\end{barticle}
\endbibitem

\bibitem{li2021epitaxial}
\begin{barticle}
\bauthor{\bsnm{Li}, \binits{T.}},
\bauthor{\bsnm{Guo}, \binits{W.}},
\bauthor{\bsnm{Ma}, \binits{L.}},
\bauthor{\bsnm{Li}, \binits{W.}},
\bauthor{\bsnm{Yu}, \binits{Z.}},
\bauthor{\bsnm{Han}, \binits{Z.}},
\bauthor{\bsnm{Gao}, \binits{S.}},
\bauthor{\bsnm{Liu}, \binits{L.}},
\bauthor{\bsnm{Fan}, \binits{D.}},
\bauthor{\bsnm{Wang}, \binits{Z.}}, \betal:
\batitle{Epitaxial growth of wafer-scale molybdenum disulfide semiconductor
  single crystals on sapphire}.
\bjtitle{Nat. Nanotechnol.}
\bvolume{16}(\bissue{11}),
\bfpage{1201}--\blpage{1207}
(\byear{2021})
\end{barticle}
\endbibitem

\bibitem{munkhbat2020transition}
\begin{barticle}
\bauthor{\bsnm{Munkhbat}, \binits{B.}},
\bauthor{\bsnm{Yankovich}, \binits{A.B.}},
\bauthor{\bsnm{Baranov}, \binits{D.G.}},
\bauthor{\bsnm{Verre}, \binits{R.}},
\bauthor{\bsnm{Olsson}, \binits{E.}},
\bauthor{\bsnm{Shegai}, \binits{T.O.}}:
\batitle{Transition metal dichalcogenide metamaterials with atomic precision}.
\bjtitle{Nat. Commun.}
\bvolume{11}(\bissue{1}),
\bfpage{1}--\blpage{8}
(\byear{2020})
\end{barticle}
\endbibitem

\bibitem{munkhbat2023nanostructured}
\begin{barticle}
\bauthor{\bsnm{Munkhbat}, \binits{B.}},
\bauthor{\bsnm{K{\"u}{\c{c}}{\"u}k{\"o}z}, \binits{B.}},
\bauthor{\bsnm{Baranov}, \binits{D.G.}},
\bauthor{\bsnm{Antosiewicz}, \binits{T.J.}},
\bauthor{\bsnm{Shegai}, \binits{T.O.}}:
\batitle{Nanostructured transition metal dichalcogenide multilayers for
  advanced nanophotonics}.
\bjtitle{Laser \& Photonics Rev.}
\bvolume{17}(\bissue{1}),
\bfpage{2200057}
(\byear{2023})
\end{barticle}
\endbibitem

\bibitem{weber2023intrinsic}
\begin{barticle}
\bauthor{\bsnm{Weber}, \binits{T.}},
\bauthor{\bsnm{K{\"u}hner}, \binits{L.}},
\bauthor{\bsnm{Sortino}, \binits{L.}},
\bauthor{\bsnm{Ben~Mhenni}, \binits{A.}},
\bauthor{\bsnm{Wilson}, \binits{N.P.}},
\bauthor{\bsnm{K{\"u}hne}, \binits{J.}},
\bauthor{\bsnm{Finley}, \binits{J.J.}},
\bauthor{\bsnm{Maier}, \binits{S.A.}},
\bauthor{\bsnm{Tittl}, \binits{A.}}:
\batitle{Intrinsic strong light-matter coupling with self-hybridized bound
  states in the continuum in van der {W}aals metasurfaces}.
\bjtitle{Nat. Mater.}
\bvolume{22},
\bfpage{970}--\blpage{976}
(\byear{2023})
\end{barticle}
\endbibitem

\bibitem{zotev2023van}
\begin{barticle}
\bauthor{\bsnm{Zotev}, \binits{P.G.}},
\bauthor{\bsnm{Wang}, \binits{Y.}},
\bauthor{\bsnm{Andres-Penares}, \binits{D.}},
\bauthor{\bsnm{Severs-Millard}, \binits{T.}},
\bauthor{\bsnm{Randerson}, \binits{S.}},
\bauthor{\bsnm{Hu}, \binits{X.}},
\bauthor{\bsnm{Sortino}, \binits{L.}},
\bauthor{\bsnm{Louca}, \binits{C.}},
\bauthor{\bsnm{Brotons-Gisbert}, \binits{M.}},
\bauthor{\bsnm{Huq}, \binits{T.}}, \betal:
\batitle{Van der waals materials for applications in nanophotonics}.
\bjtitle{Laser \& Photonics Rev.}
\bvolume{17}(\bissue{8}),
\bfpage{2200957}
(\byear{2023})
\end{barticle}
\endbibitem

\bibitem{verre2019transition}
\begin{barticle}
\bauthor{\bsnm{Verre}, \binits{R.}},
\bauthor{\bsnm{Baranov}, \binits{D.G.}},
\bauthor{\bsnm{Munkhbat}, \binits{B.}},
\bauthor{\bsnm{Cuadra}, \binits{J.}},
\bauthor{\bsnm{K{\"a}ll}, \binits{M.}},
\bauthor{\bsnm{Shegai}, \binits{T.}}:
\batitle{Transition metal dichalcogenide nanodisks as high-index dielectric
  {M}ie nanoresonators}.
\bjtitle{Nat. Nanotechnol.}
\bvolume{14}(\bissue{7}),
\bfpage{679}--\blpage{683}
(\byear{2019})
\end{barticle}
\endbibitem

\bibitem{tselikov2022transition}
\begin{barticle}
\bauthor{\bsnm{Tselikov}, \binits{G.I.}},
\bauthor{\bsnm{Ermolaev}, \binits{G.A.}},
\bauthor{\bsnm{Popov}, \binits{A.A.}},
\bauthor{\bsnm{Tikhonowski}, \binits{G.V.}},
\bauthor{\bsnm{Panova}, \binits{D.A.}},
\bauthor{\bsnm{Taradin}, \binits{A.S.}},
\bauthor{\bsnm{Vyshnevyy}, \binits{A.A.}},
\bauthor{\bsnm{Syuy}, \binits{A.V.}},
\bauthor{\bsnm{Klimentov}, \binits{S.M.}},
\bauthor{\bsnm{Novikov}, \binits{S.M.}}, \betal:
\batitle{Transition metal dichalcogenide nanospheres for high-refractive-index
  nanophotonics and biomedical theranostics}.
\bjtitle{Proc. Natl. Acad. Sci.}
\bvolume{119}(\bissue{39}),
\bfpage{2208830119}
(\byear{2022})
\end{barticle}
\endbibitem

\bibitem{maciel2023probing}
\begin{barticle}
\bauthor{\bsnm{Maciel-Escudero}, \binits{C.}},
\bauthor{\bsnm{Yankovich}, \binits{A.B.}},
\bauthor{\bsnm{Munkhbat}, \binits{B.}},
\bauthor{\bsnm{Baranov}, \binits{D.G.}},
\bauthor{\bsnm{Hillenbrand}, \binits{R.}},
\bauthor{\bsnm{Olsson}, \binits{E.}},
\bauthor{\bsnm{Aizpurua}, \binits{J.}},
\bauthor{\bsnm{Shegai}, \binits{T.O.}}:
\batitle{Probing optical anapoles with fast electron beams}.
\bjtitle{Nat. Commun.}
\bvolume{14},
\bfpage{8478}
(\byear{2023})
\end{barticle}
\endbibitem

\bibitem{zograf2023combining}
\begin{botherref}
\oauthor{\bsnm{Zograf}, \binits{G.}},
\oauthor{\bsnm{Polyakov}, \binits{A.Y.}},
\oauthor{\bsnm{Bancerek}, \binits{M.}},
\oauthor{\bsnm{Antosiewicz}, \binits{T.}},
\oauthor{\bsnm{Kucukoz}, \binits{B.}},
\oauthor{\bsnm{Shegai}, \binits{T.}}:
Combining ultrahigh index with exceptional nonlinearity in resonant transition
  metal dichalcogenide nanodisks.
arXiv preprint arXiv:2308.11504
(2023)
\end{botherref}
\endbibitem

\bibitem{sung2022room}
\begin{barticle}
\bauthor{\bsnm{Sung}, \binits{J.}},
\bauthor{\bsnm{Shin}, \binits{D.}},
\bauthor{\bsnm{Cho}, \binits{H.}},
\bauthor{\bsnm{Lee}, \binits{S.W.}},
\bauthor{\bsnm{Park}, \binits{S.}},
\bauthor{\bsnm{Kim}, \binits{Y.D.}},
\bauthor{\bsnm{Moon}, \binits{J.S.}},
\bauthor{\bsnm{Kim}, \binits{J.-H.}},
\bauthor{\bsnm{Gong}, \binits{S.-H.}}:
\batitle{Room-temperature continuous-wave indirect-bandgap transition lasing in
  an ultra-thin {WS$_2$} disk}.
\bjtitle{Nat. Photon.}
\bvolume{16},
\bfpage{792}--\blpage{797}
(\byear{2022})
\end{barticle}
\endbibitem

\bibitem{ling2021all}
\begin{barticle}
\bauthor{\bsnm{Ling}, \binits{H.}},
\bauthor{\bsnm{Li}, \binits{R.}},
\bauthor{\bsnm{Davoyan}, \binits{A.R.}}:
\batitle{All van der waals integrated nanophotonics with bulk transition metal
  dichalcogenides}.
\bjtitle{ACS Photonics}
\bvolume{8}(\bissue{3}),
\bfpage{721}--\blpage{730}
(\byear{2021})
\end{barticle}
\endbibitem

\bibitem{ling2023deeply}
\begin{barticle}
\bauthor{\bsnm{Ling}, \binits{H.}},
\bauthor{\bsnm{Manna}, \binits{A.}},
\bauthor{\bsnm{Shen}, \binits{J.}},
\bauthor{\bsnm{Tung}, \binits{H.-T.}},
\bauthor{\bsnm{Sharp}, \binits{D.}},
\bauthor{\bsnm{Fr{\"o}ch}, \binits{J.}},
\bauthor{\bsnm{Dai}, \binits{S.}},
\bauthor{\bsnm{Majumdar}, \binits{A.}},
\bauthor{\bsnm{Davoyan}, \binits{A.R.}}:
\batitle{Deeply subwavelength integrated excitonic van der waals
  nanophotonics}.
\bjtitle{Optica}
\bvolume{10}(\bissue{10}),
\bfpage{1345}--\blpage{1352}
(\byear{2023})
\end{barticle}
\endbibitem

\bibitem{li2021phase}
\begin{barticle}
\bauthor{\bsnm{Li}, \binits{W.}},
\bauthor{\bsnm{Qian}, \binits{X.}},
\bauthor{\bsnm{Li}, \binits{J.}}:
\batitle{Phase transitions in 2{D} materials}.
\bjtitle{Nat. Rev. Mater.}
\bvolume{6}(\bissue{9}),
\bfpage{829}--\blpage{846}
(\byear{2021})
\end{barticle}
\endbibitem

\bibitem{chialvo2010emergent}
\begin{barticle}
\bauthor{\bsnm{Chialvo}, \binits{D.R.}}:
\batitle{Emergent complex neural dynamics}.
\bjtitle{Nat. Phys.}
\bvolume{6}(\bissue{10}),
\bfpage{744}--\blpage{750}
(\byear{2010})
\end{barticle}
\endbibitem

\bibitem{markovic2020physics}
\begin{barticle}
\bauthor{\bsnm{Markovi{\'c}}, \binits{D.}},
\bauthor{\bsnm{Mizrahi}, \binits{A.}},
\bauthor{\bsnm{Querlioz}, \binits{D.}},
\bauthor{\bsnm{Grollier}, \binits{J.}}:
\batitle{Physics for neuromorphic computing}.
\bjtitle{Nat. Rev. Phys.}
\bvolume{2}(\bissue{9}),
\bfpage{499}--\blpage{510}
(\byear{2020})
\end{barticle}
\endbibitem

\bibitem{wuttig2007phase}
\begin{barticle}
\bauthor{\bsnm{Wuttig}, \binits{M.}},
\bauthor{\bsnm{Yamada}, \binits{N.}}:
\batitle{Phase-change materials for rewriteable data storage}.
\bjtitle{Nat. Mater.}
\bvolume{6}(\bissue{11}),
\bfpage{824}--\blpage{832}
(\byear{2007})
\end{barticle}
\endbibitem

\bibitem{seo2021reconfigurable}
\begin{barticle}
\bauthor{\bsnm{Seo}, \binits{S.-Y.}},
\bauthor{\bsnm{Moon}, \binits{G.}},
\bauthor{\bsnm{Okello}, \binits{O.F.}},
\bauthor{\bsnm{Park}, \binits{M.Y.}},
\bauthor{\bsnm{Han}, \binits{C.}},
\bauthor{\bsnm{Cha}, \binits{S.}},
\bauthor{\bsnm{Choi}, \binits{H.}},
\bauthor{\bsnm{Yeom}, \binits{H.W.}},
\bauthor{\bsnm{Choi}, \binits{S.-Y.}},
\bauthor{\bsnm{Park}, \binits{J.}}, \betal:
\batitle{Reconfigurable photo-induced doping of two-dimensional van der {W}aals
  semiconductors using different photon energies}.
\bjtitle{Nat. Electron.}
\bvolume{4}(\bissue{1}),
\bfpage{38}--\blpage{44}
(\byear{2021})
\end{barticle}
\endbibitem

\bibitem{wang2017structural}
\begin{barticle}
\bauthor{\bsnm{Wang}, \binits{Y.}},
\bauthor{\bsnm{Xiao}, \binits{J.}},
\bauthor{\bsnm{Zhu}, \binits{H.}},
\bauthor{\bsnm{Li}, \binits{Y.}},
\bauthor{\bsnm{Alsaid}, \binits{Y.}},
\bauthor{\bsnm{Fong}, \binits{K.Y.}},
\bauthor{\bsnm{Zhou}, \binits{Y.}},
\bauthor{\bsnm{Wang}, \binits{S.}},
\bauthor{\bsnm{Shi}, \binits{W.}},
\bauthor{\bsnm{Wang}, \binits{Y.}}, \betal:
\batitle{Structural phase transition in monolayer {MoTe$_2$} driven by
  electrostatic doping}.
\bjtitle{Nature}
\bvolume{550}(\bissue{7677}),
\bfpage{487}--\blpage{491}
(\byear{2017})
\end{barticle}
\endbibitem

\bibitem{voiry2013enhanced}
\begin{barticle}
\bauthor{\bsnm{Voiry}, \binits{D.}},
\bauthor{\bsnm{Yamaguchi}, \binits{H.}},
\bauthor{\bsnm{Li}, \binits{J.}},
\bauthor{\bsnm{Silva}, \binits{R.}},
\bauthor{\bsnm{Alves}, \binits{D.C.}},
\bauthor{\bsnm{Fujita}, \binits{T.}},
\bauthor{\bsnm{Chen}, \binits{M.}},
\bauthor{\bsnm{Asefa}, \binits{T.}},
\bauthor{\bsnm{Shenoy}, \binits{V.B.}},
\bauthor{\bsnm{Eda}, \binits{G.}}, \betal:
\batitle{Enhanced catalytic activity in strained chemically exfoliated {WS$_2$}
  nanosheets for hydrogen evolution}.
\bjtitle{Nat. Mater.}
\bvolume{12}(\bissue{9}),
\bfpage{850}--\blpage{855}
(\byear{2013})
\end{barticle}
\endbibitem

\bibitem{cho2015phase}
\begin{barticle}
\bauthor{\bsnm{Cho}, \binits{S.}},
\bauthor{\bsnm{Kim}, \binits{S.}},
\bauthor{\bsnm{Kim}, \binits{J.H.}},
\bauthor{\bsnm{Zhao}, \binits{J.}},
\bauthor{\bsnm{Seok}, \binits{J.}},
\bauthor{\bsnm{Keum}, \binits{D.H.}},
\bauthor{\bsnm{Baik}, \binits{J.}},
\bauthor{\bsnm{Choe}, \binits{D.-H.}},
\bauthor{\bsnm{Chang}, \binits{K.J.}},
\bauthor{\bsnm{Suenaga}, \binits{K.}}, \betal:
\batitle{Phase patterning for ohmic homojunction contact in {MoTe$_2$}}.
\bjtitle{Science}
\bvolume{349}(\bissue{6248}),
\bfpage{625}--\blpage{628}
(\byear{2015})
\end{barticle}
\endbibitem

\bibitem{guan2023femtosecond}
\begin{barticle}
\bauthor{\bsnm{Guan}, \binits{Y.}},
\bauthor{\bsnm{Ding}, \binits{Y.}},
\bauthor{\bsnm{Fang}, \binits{Y.}},
\bauthor{\bsnm{Wang}, \binits{G.}},
\bauthor{\bsnm{Zhao}, \binits{S.}},
\bauthor{\bsnm{Wang}, \binits{L.}},
\bauthor{\bsnm{Huang}, \binits{J.}},
\bauthor{\bsnm{Chen}, \binits{M.}},
\bauthor{\bsnm{Hao}, \binits{J.}},
\bauthor{\bsnm{Xu}, \binits{C.}}, \betal:
\batitle{Femtosecond laser-driven phase engineering of {WS$_2$} for
  nano-periodic phase patterning and sub-ppm ammonia gas sensing}.
\bjtitle{Small}
\bvolume{19},
\bfpage{2303654}
(\byear{2023})
\end{barticle}
\endbibitem

\bibitem{yu2018high}
\begin{barticle}
\bauthor{\bsnm{Yu}, \binits{Y.}},
\bauthor{\bsnm{Nam}, \binits{G.-H.}},
\bauthor{\bsnm{He}, \binits{Q.}},
\bauthor{\bsnm{Wu}, \binits{X.-J.}},
\bauthor{\bsnm{Zhang}, \binits{K.}},
\bauthor{\bsnm{Yang}, \binits{Z.}},
\bauthor{\bsnm{Chen}, \binits{J.}},
\bauthor{\bsnm{Ma}, \binits{Q.}},
\bauthor{\bsnm{Zhao}, \binits{M.}},
\bauthor{\bsnm{Liu}, \binits{Z.}}, \betal:
\batitle{High phase-purity {1T'}-{MoS$_2$}- and {1T'-MoSe$_2$}-layered
  crystals}.
\bjtitle{Nat. Chem.}
\bvolume{10}(\bissue{6}),
\bfpage{638}--\blpage{643}
(\byear{2018})
\end{barticle}
\endbibitem

\bibitem{nasu2004photoinduced}
\begin{bbook}
\bauthor{\bsnm{Nasu}, \binits{K.}}:
\bbtitle{Photoinduced Phase Transitions}.
\bpublisher{World Scientific},
\blocation{Singapore}
(\byear{2004})
\end{bbook}
\endbibitem

\bibitem{yang2017structural}
\begin{barticle}
\bauthor{\bsnm{Yang}, \binits{H.}},
\bauthor{\bsnm{Kim}, \binits{S.W.}},
\bauthor{\bsnm{Chhowalla}, \binits{M.}},
\bauthor{\bsnm{Lee}, \binits{Y.H.}}:
\batitle{Structural and quantum-state phase transitions in van der {W}aals
  layered materials}.
\bjtitle{Nat. Phys.}
\bvolume{13}(\bissue{10}),
\bfpage{931}--\blpage{937}
(\byear{2017})
\end{barticle}
\endbibitem

\bibitem{stojchevska2014ultrafast}
\begin{barticle}
\bauthor{\bsnm{Stojchevska}, \binits{L.}},
\bauthor{\bsnm{Vaskivskyi}, \binits{I.}},
\bauthor{\bsnm{Mertelj}, \binits{T.}},
\bauthor{\bsnm{Kusar}, \binits{P.}},
\bauthor{\bsnm{Svetin}, \binits{D.}},
\bauthor{\bsnm{Brazovskii}, \binits{S.}},
\bauthor{\bsnm{Mihailovic}, \binits{D.}}:
\batitle{Ultrafast switching to a stable hidden quantum state in an electronic
  crystal}.
\bjtitle{Science}
\bvolume{344}(\bissue{6180}),
\bfpage{177}--\blpage{180}
(\byear{2014})
\end{barticle}
\endbibitem

\bibitem{vaskivskyi2024high}
\begin{barticle}
\bauthor{\bsnm{Vaskivskyi}, \binits{I.}},
\bauthor{\bsnm{Mraz}, \binits{A.}},
\bauthor{\bsnm{Venturini}, \binits{R.}},
\bauthor{\bsnm{Jecl}, \binits{G.}},
\bauthor{\bsnm{Vaskivskyi}, \binits{Y.}},
\bauthor{\bsnm{Mincigrucci}, \binits{R.}},
\bauthor{\bsnm{Foglia}, \binits{L.}},
\bauthor{\bsnm{De~Angelis}, \binits{D.}},
\bauthor{\bsnm{Pelli-Cresi}, \binits{J.-S.}},
\bauthor{\bsnm{Paltanin}, \binits{E.}}, \betal:
\batitle{A high-efficiency programmable modulator for extreme ultraviolet light
  with nanometre feature size based on an electronic phase transition}.
\bjtitle{Nat. Photon.}
\bvolume{18},
\bfpage{458}--\blpage{463}
(\byear{2024})
\end{barticle}
\endbibitem

\bibitem{zhang2015res2}
\begin{barticle}
\bauthor{\bsnm{Zhang}, \binits{E.}},
\bauthor{\bsnm{Jin}, \binits{Y.}},
\bauthor{\bsnm{Yuan}, \binits{X.}},
\bauthor{\bsnm{Wang}, \binits{W.}},
\bauthor{\bsnm{Zhang}, \binits{C.}},
\bauthor{\bsnm{Tang}, \binits{L.}},
\bauthor{\bsnm{Liu}, \binits{S.}},
\bauthor{\bsnm{Zhou}, \binits{P.}},
\bauthor{\bsnm{Hu}, \binits{W.}},
\bauthor{\bsnm{Xiu}, \binits{F.}}:
\batitle{{ReS$_2$}-based field-effect transistors and photodetectors}.
\bjtitle{Adv. Funct. Mater.}
\bvolume{25}(\bissue{26}),
\bfpage{4076}--\blpage{4082}
(\byear{2015})
\end{barticle}
\endbibitem

\bibitem{pradhan2015metal}
\begin{barticle}
\bauthor{\bsnm{Pradhan}, \binits{N.R.}},
\bauthor{\bsnm{McCreary}, \binits{A.}},
\bauthor{\bsnm{Rhodes}, \binits{D.}},
\bauthor{\bsnm{Lu}, \binits{Z.}},
\bauthor{\bsnm{Feng}, \binits{S.}},
\bauthor{\bsnm{Manousakis}, \binits{E.}},
\bauthor{\bsnm{Smirnov}, \binits{D.}},
\bauthor{\bsnm{Namburu}, \binits{R.}},
\bauthor{\bsnm{Dubey}, \binits{M.}},
\bauthor{\bsnm{Hight~Walker}, \binits{A.R.}}, \betal:
\batitle{Metal to insulator quantum-phase transition in few-layered {ReS$_2$}}.
\bjtitle{Nano Lett.}
\bvolume{15}(\bissue{12}),
\bfpage{8377}--\blpage{8384}
(\byear{2015})
\end{barticle}
\endbibitem

\bibitem{rahman2017advent}
\begin{barticle}
\bauthor{\bsnm{Rahman}, \binits{M.}},
\bauthor{\bsnm{Davey}, \binits{K.}},
\bauthor{\bsnm{Qiao}, \binits{S.-Z.}}:
\batitle{Advent of 2{D} rhenium disulfide ({ReS$_2$}): fundamentals to
  applications}.
\bjtitle{Adv. Funct. Mater.}
\bvolume{27}(\bissue{10}),
\bfpage{1606129}
(\byear{2017})
\end{barticle}
\endbibitem

\bibitem{xiang2019anomalous}
\begin{barticle}
\bauthor{\bsnm{Xiang}, \binits{D.}},
\bauthor{\bsnm{Liu}, \binits{T.}},
\bauthor{\bsnm{Wang}, \binits{J.}},
\bauthor{\bsnm{Wang}, \binits{P.}},
\bauthor{\bsnm{Wang}, \binits{L.}},
\bauthor{\bsnm{Zheng}, \binits{Y.}},
\bauthor{\bsnm{Wang}, \binits{Y.}},
\bauthor{\bsnm{Gao}, \binits{J.}},
\bauthor{\bsnm{Ang}, \binits{K.-W.}},
\bauthor{\bsnm{Eda}, \binits{G.}}, \betal:
\batitle{Anomalous broadband spectrum photodetection in 2{D} rhenium disulfide
  transistor}.
\bjtitle{Adv. Opt. Mater.}
\bvolume{7}(\bissue{23}),
\bfpage{1901115}
(\byear{2019})
\end{barticle}
\endbibitem

\bibitem{gogna2020self}
\begin{barticle}
\bauthor{\bsnm{Gogna}, \binits{R.}},
\bauthor{\bsnm{Zhang}, \binits{L.}},
\bauthor{\bsnm{Deng}, \binits{H.}}:
\batitle{Self-hybridized, polarized polaritons in {ReS$_2$} crystals}.
\bjtitle{ACS Photonics}
\bvolume{7}(\bissue{12}),
\bfpage{3328}--\blpage{3332}
(\byear{2020})
\end{barticle}
\endbibitem

\bibitem{xiong2022nonvolatile}
\begin{barticle}
\bauthor{\bsnm{Xiong}, \binits{X.}},
\bauthor{\bsnm{Kang}, \binits{J.}},
\bauthor{\bsnm{Liu}, \binits{S.}},
\bauthor{\bsnm{Tong}, \binits{A.}},
\bauthor{\bsnm{Fu}, \binits{T.}},
\bauthor{\bsnm{Li}, \binits{X.}},
\bauthor{\bsnm{Huang}, \binits{R.}},
\bauthor{\bsnm{Wu}, \binits{Y.}}:
\batitle{Nonvolatile logic and ternary content-addressable memory based on
  complementary black phosphorus and rhenium disulfide transistors}.
\bjtitle{Adv. Mater.}
\bvolume{34}(\bissue{48}),
\bfpage{2106321}
(\byear{2022})
\end{barticle}
\endbibitem

\bibitem{tongay2014monolayer}
\begin{barticle}
\bauthor{\bsnm{Tongay}, \binits{S.}},
\bauthor{\bsnm{Sahin}, \binits{H.}},
\bauthor{\bsnm{Ko}, \binits{C.}},
\bauthor{\bsnm{Luce}, \binits{A.}},
\bauthor{\bsnm{Fan}, \binits{W.}},
\bauthor{\bsnm{Liu}, \binits{K.}},
\bauthor{\bsnm{Zhou}, \binits{J.}},
\bauthor{\bsnm{Huang}, \binits{Y.-S.}},
\bauthor{\bsnm{Ho}, \binits{C.-H.}},
\bauthor{\bsnm{Yan}, \binits{J.}}, \betal:
\batitle{Monolayer behaviour in bulk {ReS$_2$} due to electronic and
  vibrational decoupling}.
\bjtitle{Nat. Commun.}
\bvolume{5},
\bfpage{3252}
(\byear{2014})
\end{barticle}
\endbibitem

\bibitem{lin2015single}
\begin{barticle}
\bauthor{\bsnm{Lin}, \binits{Y.-C.}},
\bauthor{\bsnm{Komsa}, \binits{H.-P.}},
\bauthor{\bsnm{Yeh}, \binits{C.-H.}},
\bauthor{\bsnm{Bjorkman}, \binits{T.}},
\bauthor{\bsnm{Liang}, \binits{Z.-Y.}},
\bauthor{\bsnm{Ho}, \binits{C.-H.}},
\bauthor{\bsnm{Huang}, \binits{Y.-S.}},
\bauthor{\bsnm{Chiu}, \binits{P.-W.}},
\bauthor{\bsnm{Krasheninnikov}, \binits{A.V.}},
\bauthor{\bsnm{Suenaga}, \binits{K.}}:
\batitle{Single-layer {ReS$_2$}: two-dimensional semiconductor with tunable
  in-plane anisotropy}.
\bjtitle{ACS Nano}
\bvolume{9}(\bissue{11}),
\bfpage{11249}--\blpage{11257}
(\byear{2015})
\end{barticle}
\endbibitem

\bibitem{chenet2015plane}
\begin{barticle}
\bauthor{\bsnm{Chenet}, \binits{D.A.}},
\bauthor{\bsnm{Aslan}, \binits{B.}},
\bauthor{\bsnm{Huang}, \binits{P.Y.}},
\bauthor{\bsnm{Fan}, \binits{C.}},
\bauthor{\bsnm{Van Der~Zande}, \binits{A.M.}},
\bauthor{\bsnm{Heinz}, \binits{T.F.}},
\bauthor{\bsnm{Hone}, \binits{J.C.}}:
\batitle{In-plane anisotropy in mono-and few-layer {ReS$_2$} probed by raman
  spectroscopy and scanning transmission electron microscopy}.
\bjtitle{Nano Lett.}
\bvolume{15}(\bissue{9}),
\bfpage{5667}--\blpage{5672}
(\byear{2015})
\end{barticle}
\endbibitem

\bibitem{liu2016highly}
\begin{barticle}
\bauthor{\bsnm{Liu}, \binits{F.}},
\bauthor{\bsnm{Zheng}, \binits{S.}},
\bauthor{\bsnm{He}, \binits{X.}},
\bauthor{\bsnm{Chaturvedi}, \binits{A.}},
\bauthor{\bsnm{He}, \binits{J.}},
\bauthor{\bsnm{Chow}, \binits{W.L.}},
\bauthor{\bsnm{Mion}, \binits{T.R.}},
\bauthor{\bsnm{Wang}, \binits{X.}},
\bauthor{\bsnm{Zhou}, \binits{J.}},
\bauthor{\bsnm{Fu}, \binits{Q.}}, \betal:
\batitle{Highly sensitive detection of polarized light using anisotropic 2{D}
  {ReS$_2$}}.
\bjtitle{Adv. Funct. Mater.}
\bvolume{26}(\bissue{8}),
\bfpage{1169}--\blpage{1177}
(\byear{2016})
\end{barticle}
\endbibitem

\bibitem{ermolaev2024wandering}
\begin{barticle}
\bauthor{\bsnm{Ermolaev}, \binits{G.A.}},
\bauthor{\bsnm{Voronin}, \binits{K.V.}},
\bauthor{\bsnm{Toksumakov}, \binits{A.N.}},
\bauthor{\bsnm{Grudinin}, \binits{D.V.}},
\bauthor{\bsnm{Fradkin}, \binits{I.M.}},
\bauthor{\bsnm{Mazitov}, \binits{A.}},
\bauthor{\bsnm{Slavich}, \binits{A.S.}},
\bauthor{\bsnm{Tatmyshevskiy}, \binits{M.K.}},
\bauthor{\bsnm{Yakubovsky}, \binits{D.I.}},
\bauthor{\bsnm{Solovey}, \binits{V.R.}}, \betal:
\batitle{Wandering principal optical axes in van der {W}aals triclinic
  materials}.
\bjtitle{Nat. Commun.}
\bvolume{15}(\bissue{1}),
\bfpage{1552}
(\byear{2024})
\end{barticle}
\endbibitem

\bibitem{zhou2020stacking}
\begin{barticle}
\bauthor{\bsnm{Zhou}, \binits{Y.}},
\bauthor{\bsnm{Maity}, \binits{N.}},
\bauthor{\bsnm{Rai}, \binits{A.}},
\bauthor{\bsnm{Juneja}, \binits{R.}},
\bauthor{\bsnm{Meng}, \binits{X.}},
\bauthor{\bsnm{Roy}, \binits{A.}},
\bauthor{\bsnm{Zhang}, \binits{Y.}},
\bauthor{\bsnm{Xu}, \binits{X.}},
\bauthor{\bsnm{Lin}, \binits{J.-F.}},
\bauthor{\bsnm{Banerjee}, \binits{S.K.}}, \betal:
\batitle{Stacking-order-driven optical properties and carrier dynamics in
  {ReS$_2$}}.
\bjtitle{Adv. Mater.}
\bvolume{32}(\bissue{22}),
\bfpage{1908311}
(\byear{2020})
\end{barticle}
\endbibitem

\bibitem{song2018extraordinary}
\begin{barticle}
\bauthor{\bsnm{Song}, \binits{Y.}},
\bauthor{\bsnm{Hu}, \binits{S.}},
\bauthor{\bsnm{Lin}, \binits{M.-L.}},
\bauthor{\bsnm{Gan}, \binits{X.}},
\bauthor{\bsnm{Tan}, \binits{P.-H.}},
\bauthor{\bsnm{Zhao}, \binits{J.}}:
\batitle{Extraordinary second harmonic generation in {ReS$_2$} atomic
  crystals}.
\bjtitle{ACS Photonics}
\bvolume{5}(\bissue{9}),
\bfpage{3485}--\blpage{3491}
(\byear{2018})
\end{barticle}
\endbibitem

\bibitem{kucukoz2022boosting}
\begin{barticle}
\bauthor{\bsnm{K{\"u}{\c{c}}{\"u}k{\"o}z}, \binits{B.}},
\bauthor{\bsnm{Munkhbat}, \binits{B.}},
\bauthor{\bsnm{Shegai}, \binits{T.O.}}:
\batitle{Boosting second-harmonic generation in monolayer rhenium disulfide by
  reversible laser patterning}.
\bjtitle{ACS Photonics}
\bvolume{9}(\bissue{2}),
\bfpage{518}--\blpage{526}
(\byear{2022})
\end{barticle}
\endbibitem

\bibitem{castellanos2010optical}
\begin{botherref}
\oauthor{\bsnm{Castellanos-Gomez}, \binits{A.}},
\oauthor{\bsnm{Agra{\"\i}t}, \binits{N.}},
\oauthor{\bsnm{Rubio-Bollinger}, \binits{G.}}:
Optical identification of atomically thin dichalcogenide crystals.
Appl. Phys. Lett.
\textbf{96}(21)
(2010)
\end{botherref}
\endbibitem

\bibitem{anzai2019broad}
\begin{barticle}
\bauthor{\bsnm{Anzai}, \binits{Y.}},
\bauthor{\bsnm{Yamamoto}, \binits{M.}},
\bauthor{\bsnm{Genchi}, \binits{S.}},
\bauthor{\bsnm{Watanabe}, \binits{K.}},
\bauthor{\bsnm{Taniguchi}, \binits{T.}},
\bauthor{\bsnm{Ichikawa}, \binits{S.}},
\bauthor{\bsnm{Fujiwara}, \binits{Y.}},
\bauthor{\bsnm{Tanaka}, \binits{H.}}:
\batitle{Broad range thickness identification of hexagonal boron nitride by
  colors}.
\bjtitle{Appl. Phys. Express}
\bvolume{12}(\bissue{5}),
\bfpage{055007}
(\byear{2019})
\end{barticle}
\endbibitem

\bibitem{lebeau2010position}
\begin{barticle}
\bauthor{\bsnm{LeBeau}, \binits{J.M.}},
\bauthor{\bsnm{Findlay}, \binits{S.D.}},
\bauthor{\bsnm{Allen}, \binits{L.J.}},
\bauthor{\bsnm{Stemmer}, \binits{S.}}:
\batitle{Position averaged convergent beam electron diffraction: Theory and
  applications}.
\bjtitle{Ultramicroscopy}
\bvolume{110}(\bissue{2}),
\bfpage{118}--\blpage{125}
(\byear{2010})
\end{barticle}
\endbibitem

\bibitem{ophus2019four}
\begin{barticle}
\bauthor{\bsnm{Ophus}, \binits{C.}}:
\batitle{Four-dimensional scanning transmission electron microscopy
  ({4D}-{STEM}): From scanning nanodiffraction to ptychography and beyond}.
\bjtitle{Microscopy and Microanalysis}
\bvolume{25}(\bissue{3}),
\bfpage{563}--\blpage{582}
(\byear{2019})
\end{barticle}
\endbibitem

\bibitem{zhang2020atomic}
\begin{barticle}
\bauthor{\bsnm{Zhang}, \binits{C.}},
\bauthor{\bsnm{Feng}, \binits{J.}},
\bauthor{\bsnm{DaCosta}, \binits{L.R.}},
\bauthor{\bsnm{Voyles}, \binits{P.M.}}:
\batitle{Atomic resolution convergent beam electron diffraction analysis using
  convolutional neural networks}.
\bjtitle{Ultramicroscopy}
\bvolume{210},
\bfpage{112921}
(\byear{2020})
\end{barticle}
\endbibitem

\bibitem{komsa2012two}
\begin{barticle}
\bauthor{\bsnm{Komsa}, \binits{H.-P.}},
\bauthor{\bsnm{Kotakoski}, \binits{J.}},
\bauthor{\bsnm{Kurasch}, \binits{S.}},
\bauthor{\bsnm{Lehtinen}, \binits{O.}},
\bauthor{\bsnm{Kaiser}, \binits{U.}},
\bauthor{\bsnm{Krasheninnikov}, \binits{A.V.}}:
\batitle{Two-dimensional transition metal dichalcogenides under electron
  irradiation: defect production and doping}.
\bjtitle{Phys. Rev. Lett.}
\bvolume{109}(\bissue{3}),
\bfpage{035503}
(\byear{2012})
\end{barticle}
\endbibitem

\bibitem{koster2023phase}
\begin{barticle}
\bauthor{\bsnm{K{\"o}ster}, \binits{J.}},
\bauthor{\bsnm{Kretschmer}, \binits{S.}},
\bauthor{\bsnm{Storm}, \binits{A.}},
\bauthor{\bsnm{Rasper}, \binits{F.}},
\bauthor{\bsnm{Kinyanjui}, \binits{M.K.}},
\bauthor{\bsnm{Krasheninnikov}, \binits{A.V.}},
\bauthor{\bsnm{Kaiser}, \binits{U.}}:
\batitle{Phase transformations in single-layer {MoTe$_2$} stimulated by
  electron irradiation and annealing}.
\bjtitle{Nanotechnology}
\bvolume{35},
\bfpage{145301}
(\byear{2024})
\end{barticle}
\endbibitem

\bibitem{qiao2016polytypism}
\begin{barticle}
\bauthor{\bsnm{Qiao}, \binits{X.-F.}},
\bauthor{\bsnm{Wu}, \binits{J.-B.}},
\bauthor{\bsnm{Zhou}, \binits{L.}},
\bauthor{\bsnm{Qiao}, \binits{J.}},
\bauthor{\bsnm{Shi}, \binits{W.}},
\bauthor{\bsnm{Chen}, \binits{T.}},
\bauthor{\bsnm{Zhang}, \binits{X.}},
\bauthor{\bsnm{Zhang}, \binits{J.}},
\bauthor{\bsnm{Ji}, \binits{W.}},
\bauthor{\bsnm{Tan}, \binits{P.-H.}}:
\batitle{Polytypism and unexpected strong interlayer coupling in
  two-dimensional layered {ReS$_2$}}.
\bjtitle{Nanoscale}
\bvolume{8}(\bissue{15}),
\bfpage{8324}--\blpage{8332}
(\byear{2016})
\end{barticle}
\endbibitem

\bibitem{huang2018enhanced}
\begin{barticle}
\bauthor{\bsnm{Huang}, \binits{J.}},
\bauthor{\bsnm{Gao}, \binits{H.}},
\bauthor{\bsnm{Xia}, \binits{Y.}},
\bauthor{\bsnm{Sun}, \binits{Y.}},
\bauthor{\bsnm{Xiong}, \binits{J.}},
\bauthor{\bsnm{Li}, \binits{Y.}},
\bauthor{\bsnm{Cong}, \binits{S.}},
\bauthor{\bsnm{Guo}, \binits{J.}},
\bauthor{\bsnm{Du}, \binits{S.}},
\bauthor{\bsnm{Zou}, \binits{G.}}:
\batitle{Enhanced photoelectrochemical performance of defect-rich {ReS$_2$}
  nanosheets in visible-light assisted hydrogen generation}.
\bjtitle{Nano Energy}
\bvolume{46},
\bfpage{305}--\blpage{313}
(\byear{2018})
\end{barticle}
\endbibitem

\bibitem{horzum2014formation}
\begin{barticle}
\bauthor{\bsnm{Horzum}, \binits{S.}},
\bauthor{\bsnm{{\c{C}}ak{\i}r}, \binits{D.}},
\bauthor{\bsnm{Suh}, \binits{J.}},
\bauthor{\bsnm{Tongay}, \binits{S.}},
\bauthor{\bsnm{Huang}, \binits{Y.-S.}},
\bauthor{\bsnm{Ho}, \binits{C.-H.}},
\bauthor{\bsnm{Wu}, \binits{J.}},
\bauthor{\bsnm{Sahin}, \binits{H.}},
\bauthor{\bsnm{Peeters}, \binits{F.}}:
\batitle{Formation and stability of point defects in monolayer rhenium
  disulfide}.
\bjtitle{Phys. Rev. B}
\bvolume{89}(\bissue{15}),
\bfpage{155433}
(\byear{2014})
\end{barticle}
\endbibitem

\bibitem{castellanos2014deterministic}
\begin{barticle}
\bauthor{\bsnm{Castellanos-Gomez}, \binits{A.}},
\bauthor{\bsnm{Buscema}, \binits{M.}},
\bauthor{\bsnm{Molenaar}, \binits{R.}},
\bauthor{\bsnm{Singh}, \binits{V.}},
\bauthor{\bsnm{Janssen}, \binits{L.}},
\bauthor{\bsnm{Van Der~Zant}, \binits{H.S.}},
\bauthor{\bsnm{Steele}, \binits{G.A.}}:
\batitle{Deterministic transfer of two-dimensional materials by all-dry
  viscoelastic stamping}.
\bjtitle{2D Materials}
\bvolume{1}(\bissue{1}),
\bfpage{011002}
(\byear{2014})
\end{barticle}
\endbibitem

\bibitem{yankovich2014picometre}
\begin{barticle}
\bauthor{\bsnm{Yankovich}, \binits{A.B.}},
\bauthor{\bsnm{Berkels}, \binits{B.}},
\bauthor{\bsnm{Dahmen}, \binits{W.}},
\bauthor{\bsnm{Binev}, \binits{P.}},
\bauthor{\bsnm{Sanchez}, \binits{S.I.}},
\bauthor{\bsnm{Bradley}, \binits{S.A.}},
\bauthor{\bsnm{Li}, \binits{A.}},
\bauthor{\bsnm{Szlufarska}, \binits{I.}},
\bauthor{\bsnm{Voyles}, \binits{P.M.}}:
\batitle{Picometre-precision analysis of scanning transmission electron
  microscopy images of platinum nanocatalysts}.
\bjtitle{Nat. Commun.}
\bvolume{5},
\bfpage{4155}
(\byear{2014})
\end{barticle}
\endbibitem

\bibitem{dacosta2021prismatic}
\begin{barticle}
\bauthor{\bsnm{DaCosta}, \binits{L.R.}},
\bauthor{\bsnm{Brown}, \binits{H.G.}},
\bauthor{\bsnm{Pelz}, \binits{P.M.}},
\bauthor{\bsnm{Rakowski}, \binits{A.}},
\bauthor{\bsnm{Barber}, \binits{N.}},
\bauthor{\bsnm{O’Donovan}, \binits{P.}},
\bauthor{\bsnm{McBean}, \binits{P.}},
\bauthor{\bsnm{Jones}, \binits{L.}},
\bauthor{\bsnm{Ciston}, \binits{J.}},
\bauthor{\bsnm{Scott}, \binits{M.}}, \betal:
\batitle{Prismatic 2.0--simulation software for scanning and high resolution
  transmission electron microscopy ({STEM} and {HRTEM})}.
\bjtitle{Micron}
\bvolume{151},
\bfpage{103141}
(\byear{2021})
\end{barticle}
\endbibitem

\bibitem{Blo94}
\begin{barticle}
\bauthor{\bsnm{Bl\"ochl}, \binits{P.E.}}:
\batitle{Projector augmented-wave method}.
\bjtitle{Phys. Rev. B}
\bvolume{50},
\bfpage{17953}--\blpage{17979}
(\byear{1994})
\end{barticle}
\endbibitem

\bibitem{KreHaf93}
\begin{barticle}
\bauthor{\bsnm{Kresse}, \binits{G.}},
\bauthor{\bsnm{Hafner}, \binits{J.}}:
\batitle{\textit{Ab initio} molecular dynamics for liquid metals}.
\bjtitle{Phys. Rev. B}
\bvolume{47},
\bfpage{558}--\blpage{561}
(\byear{1993})
\end{barticle}
\endbibitem

\bibitem{KreFur96}
\begin{barticle}
\bauthor{\bsnm{Kresse}, \binits{G.}},
\bauthor{\bsnm{Furthm\"uller}, \binits{J.}}:
\batitle{Efficiency of \textit{ab-initio} total energy calculations for metals
  and semiconductors using a plane-wave basis set}.
\bjtitle{Comput. Mater. Sci.}
\bvolume{6}(\bissue{1}),
\bfpage{15}--\blpage{50}
(\byear{1996})
\end{barticle}
\endbibitem

\bibitem{DioRydSch04}
\begin{barticle}
\bauthor{\bsnm{Dion}, \binits{M.}},
\bauthor{\bsnm{Rydberg}, \binits{H.}},
\bauthor{\bsnm{Schr\"oder}, \binits{E.}},
\bauthor{\bsnm{Langreth}, \binits{D.C.}},
\bauthor{\bsnm{Lundqvist}, \binits{B.I.}}:
\batitle{Van der {W}aals density functional for general geometries}.
\bjtitle{Phys. Rev. Lett.}
\bvolume{92},
\bfpage{246401}
(\byear{2004})
\end{barticle}
\endbibitem

\bibitem{BerHyl2014}
\begin{barticle}
\bauthor{\bsnm{Berland}, \binits{K.}},
\bauthor{\bsnm{Hyldgaard}, \binits{P.}}:
\batitle{Exchange functional that tests the robustness of the plasmon
  description of the van der {W}aals density functional}.
\bjtitle{Phys. Rev. B}
\bvolume{89},
\bfpage{035412}
(\byear{2014})
\end{barticle}
\endbibitem

\bibitem{LinErh16}
\begin{botherref}
\oauthor{\bsnm{Lindroth}, \binits{D.O.}},
\oauthor{\bsnm{Erhart}, \binits{P.}}:
Thermal transport in van der {W}aals solids from first-principles calculations.
Phys. Rev. B
\textbf{94}(11)
(2016)
\end{botherref}
\endbibitem

\bibitem{EriFraLin23}
\begin{barticle}
\bauthor{\bsnm{Eriksson}, \binits{F.}},
\bauthor{\bsnm{Fransson}, \binits{E.}},
\bauthor{\bsnm{Linderälv}, \binits{C.}},
\bauthor{\bsnm{Fan}, \binits{Z.}},
\bauthor{\bsnm{Erhart}, \binits{P.}}:
\batitle{Tuning the lattice thermal conductivity in van-der-{W}aals structures
  through rotational (dis)ordering}.
\bjtitle{ACS Nano}
\bvolume{17},
\bfpage{25565}
(\byear{2023})
\end{barticle}
\endbibitem

\bibitem{HedLun70}
\begin{bchapter}
\bauthor{\bsnm{Hedin}, \binits{L.}},
\bauthor{\bsnm{Lundqvist}, \binits{S.}}:
\bctitle{Effects of electron-electron and electron-phonon interactions on the
  one-electron states of solids}.
In: \beditor{\bsnm{Frederick~Seitz}, \binits{D.T.}},
\beditor{\bsnm{Ehrenreich}, \binits{H.}} (eds.)
\bbtitle{Solid State Phys.}
vol. \bseriesno{23},
pp. \bfpage{1}--\blpage{181}
(\byear{1970})
\end{bchapter}
\endbibitem

\bibitem{TogTan15}
\begin{barticle}
\bauthor{\bsnm{Togo}, \binits{A.}},
\bauthor{\bsnm{Tanaka}, \binits{I.}}:
\batitle{First principles phonon calculations in materials science}.
\bjtitle{Scr. Mater.}
\bvolume{108},
\bfpage{1}--\blpage{5}
(\byear{2015})
\end{barticle}
\endbibitem

\bibitem{EriFraErh19}
\begin{barticle}
\bauthor{\bsnm{Eriksson}, \binits{F.}},
\bauthor{\bsnm{Fransson}, \binits{E.}},
\bauthor{\bsnm{Erhart}, \binits{P.}}:
\batitle{The hiphive package for the extraction of high-order force constants
  by machine learning}.
\bjtitle{Adv. Theory Simul.}
\bvolume{2},
\bfpage{1800184}
(\byear{2019})
\end{barticle}
\endbibitem

\bibitem{Taghizadeh2021}
\begin{barticle}
\bauthor{\bsnm{Taghizadeh}, \binits{A.}},
\bauthor{\bsnm{Thygesen}, \binits{K.S.}},
\bauthor{\bsnm{Pedersen}, \binits{T.G.}}:
\batitle{Two-dimensional materials with giant optical nonlinearities near the
  theoretical upper limit}.
\bjtitle{ACS Nano}
\bvolume{14}(\bissue{4}),
\bfpage{7155}--\blpage{7167}
(\byear{2021})
\end{barticle}
\endbibitem

\bibitem{2005_PRB_71_035109_mortensen}
\begin{barticle}
\bauthor{\bsnm{Mortensen}, \binits{J.J.}},
\bauthor{\bsnm{Hansen}, \binits{L.B.}},
\bauthor{\bsnm{Jacobsen}, \binits{K.W.}}:
\batitle{Real-space grid implementation of the projector augmented wave
  method}.
\bjtitle{Phys. Rev. B}
\bvolume{71},
\bfpage{035109}
(\byear{2005})
\end{barticle}
\endbibitem

\bibitem{2010_JPhysCondMat_22_253202_enkovaara}
\begin{barticle}
\bauthor{\bsnm{Enkovaara}, \binits{J.}},
\bauthor{\bsnm{Rostgaard}, \binits{C.}},
\bauthor{\bsnm{Mortensen}, \binits{J.J.}},
\bauthor{\bsnm{Chen}, \binits{J.}},
\bauthor{\bsnm{Dulak}, \binits{M.}},
\bauthor{\bsnm{Ferrighi}, \binits{L.}},
\bauthor{\bparticle{an} \bsnm{C.~Glinsvad}, \binits{J.G.}},
\bauthor{\bsnm{Haikola}, \binits{V.}},
\bauthor{\bsnm{Hansen}, \binits{H.A.}},
\bauthor{\bsnm{Kristoffersen}, \binits{H.H.}},
\bauthor{\bsnm{Kuisma}, \binits{M.}},
\bauthor{\bsnm{Larsen}, \binits{A.H.}},
\bauthor{\bsnm{Lehtovaara}, \binits{L.}},
\bauthor{\bsnm{Ljungberg}, \binits{M.}},
\bauthor{\bsnm{Lopez-Acevedo}, \binits{O.}},
\bauthor{\bsnm{Moses}, \binits{P.G.}},
\bauthor{\bsnm{Ojanen}, \binits{J.}},
\bauthor{\bsnm{Olsen}, \binits{T.}},
\bauthor{\bsnm{Petzold}, \binits{V.}},
\bauthor{\bsnm{Romero}, \binits{N.A.}},
\bauthor{\bsnm{Stausholm-Moller}, \binits{J.}},
\bauthor{\bsnm{Strange}, \binits{M.}},
\bauthor{\bsnm{Tritsaris}, \binits{G.A.}},
\bauthor{\bsnm{Vanin}, \binits{M.}},
\bauthor{\bsnm{Walter}, \binits{M.}},
\bauthor{\bsnm{Hammer}, \binits{B.}},
\bauthor{\bsnm{Hakkinen}, \binits{H.}},
\bauthor{\bsnm{Madsen}, \binits{G.K.H.}},
\bauthor{\bsnm{Mieminen}, \binits{R.M.}},
\bauthor{\bsnm{Norskov}, \binits{J.K.}},
\bauthor{\bsnm{Puska}, \binits{M.}},
\bauthor{\bsnm{Rantala}, \binits{T.T.}},
\bauthor{\bsnm{Schiotz}, \binits{J.}},
\bauthor{\bsnm{Thygesen}, \binits{K.S.}},
\bauthor{\bsnm{Jacobsen}, \binits{K.W.}}:
\batitle{Electronic structure calculations with {GPAW}: a real-space
  implementation of the projector augmented-wave method}.
\bjtitle{J. Phys.: Condens. Matter}
\bvolume{22},
\bfpage{253202}
(\byear{2010})
\end{barticle}
\endbibitem

\bibitem{1996_PRL_77_3865_perdew}
\begin{barticle}
\bauthor{\bsnm{Perdew}, \binits{J.P.}},
\bauthor{\bsnm{Burke}, \binits{K.}},
\bauthor{\bsnm{Ernzerhof}, \binits{M.}}:
\batitle{Generalized gradient approximation made simple}.
\bjtitle{Phys. Rev. Lett.}
\bvolume{77},
\bfpage{3865}--\blpage{3868}
(\byear{1996})
\end{barticle}
\endbibitem

\end{thebibliography}
\end{document}